\PassOptionsToPackage{table}{xcolor}
\documentclass[twocolumn,trackchanges]{aastex701}

\makeatletter
\AtBeginDocument{\long\def\frontmatter@title@above{}}
\makeatother

\pdfoutput=1

\usepackage{epsfig}
\usepackage{graphics}
\usepackage{graphicx}
\usepackage{amsmath,amsfonts,amssymb,amsthm}
\usepackage[normalem]{ulem}
\usepackage{alltt}
\usepackage{array}
\usepackage{tabularx,tabulary}
\newcommand{\pdesc}[1]{\parbox[t]{3.4cm}{\raggedright #1}}
\usepackage[varg]{txfonts}
\usepackage{float}
\usepackage{dcolumn}
\usepackage[nolist,nohyperlinks]{acronym}
\usepackage{xspace}
\usepackage[english]{babel}
\usepackage[abs]{overpic}
\usepackage{pict2e}
\usepackage{subcaption}
\allowdisplaybreaks[1]
\usepackage[utf8]{inputenc}
\usepackage{gensymb}
\usepackage{bm}
\usepackage{stackengine}
\usepackage{multirow}
\usepackage{braket}

\usepackage{rotating}
\usepackage{booktabs}
\usepackage{longtable}

\usepackage{adjustbox}

\graphicspath{{}}
\newcommand{\AEI}{Max Planck Institute for Gravitational Physics (Albert Einstein Institute), Am M\"uhlenberg 1, Potsdam 14476, Germany}
\newcommand{\Maryland}{Department of Physics, University of Maryland, College Park, MD 20742, USA}
\newcommand{\MPIIS}{Max Planck Institute for Intelligent Systems, Max-Planck-Ring 4, 72076 T\"ubingen, Germany}
\newcommand{\MLS}{Machine Learning in Science, University of T\"ubingen \& T\"ubingen AI Center, 72076 T\"ubingen, Germany}

\newcommand{\UoN}{Nottingham Centre of Gravity \& School of Mathematical Sciences, University of Nottingham, University Park, Nottingham NG7 2RD, United Kingdom}

\newcommand{\UBC}{Department of Physics \& Astronomy, University of British Columbia, Vancouver, BC V6T 1Z1, Canada}
\newcommand{\CaltechTAPIR}{TAPIR, California Institute of Technology, Pasadena, CA 91125, USA}
\newcommand{\CaltechLIGO}{LIGO Laboratory, California Institute of Technology, Pasadena, CA 91125, USA}
\newcommand{\UIB}{Departament de F\'isica, Universitat de les Illes Balears, IAC3 -- IEEC, Crta. Valldemossa km 7.5, E-07122 Palma, Spain}
\newcommand{\ELLIS}{ELLIS Institute T\"ubingen, Maria-von-Linden-Straße 2, 72076 T\"ubingen, Germany}
\newcommand{\TUEAI}{T\"ubingen AI Center, Maria-von-Linden-Straße 1, 72076 T\"ubingen, Germany}
\newcommand{\Bicocca}{Dipartimento di Fisica ``G. Occhialini'', Universit\'a degli Studi di Milano-Bicocca, Piazza della Scienza 3, Milano, 20126, Italy}
\newcommand{\INFNMiB}{INFN, Sezione di Milano-Bicocca, Universit\'a degli Studi di Milano-Bicocca, Milano, 20126, Italy}

\newcommand{\ColorOne}{\cellcolor[HTML]{FABCA4}} 
\newcommand{\ColorTwo}{\cellcolor[HTML]{FED7B0}} 
\newcommand{\ColorThree}{\cellcolor[HTML]{FFFAD6}} 
\newcommand{\ColorFour}{\cellcolor[HTML]{E6F4C1}} 
\newcommand{\ColorFive}{\cellcolor[HTML]{C8E7B3}} 
\newcommand{\ColorSix}{\cellcolor[HTML]{ABDAAF}} 
\newcommand{\ColorSeven}{\cellcolor[HTML]{92CEA9}} 
\newcommand{\ColorNine}{\cellcolor[HTML]{84BB9F}} 
\newcommand{\ColorTen}{\cellcolor[HTML]{79B597}} 

\begin{document}

\title{Eccentricity constraints disfavor single-single capture in nuclear star clusters as the origin of all\\ LIGO-Virgo-KAGRA binary black holes}


\correspondingauthor{Nihar Gupte}
\author[0000-0002-7287-5151]{Nihar Gupte}
\email{nihar.gupte@aei.mpg.de}
\affiliation{\AEI}
\affiliation{\Maryland}

\author[0000-0002-2666-728X]{M. Coleman Miller}
\email{mcmiller@umd.edu}
\affiliation{\Maryland}

\author[0000-0001-6877-3278]{Rhiannon Udall}
\email{rhiannon.udall@ubc.ca}
\affiliation{\UBC}
\affiliation{\CaltechTAPIR}
\affiliation{\CaltechLIGO}

\author[0000-0002-0267-3562]{Sophie Bini}
\email{bini@caltech.edu}
\affiliation{\CaltechLIGO}

\author[0000-0002-2685-1538]{Alexandre Toubiana}
\email{alexandre.toubiana@unimib.it}
\affiliation{\Bicocca}
\affiliation{\INFNMiB}

\author[0000-0002-5433-1409]{Alessandra Buonanno}
\email{alessandra.buonanno@aei.mpg.de}
\affiliation{\AEI}
\affiliation{\Maryland}

\author[0000-0002-1671-3668]{Jonathan Gair}
\email{jonathan.gair@aei.mpg.de}
\affiliation{\AEI}

\author[0000-0001-8391-5596]{Aldo Gamboa}
\email{aldo.gamboa@aei.mpg.de}
\affiliation{\AEI} 

\author[0000-0002-0710-6778]{Lorenzo Pompili}
\email{lorenzo.pompili@nottingham.ac.uk}
\affiliation{\UoN}
\affiliation{\AEI}

\author[0000-0002-6874-7421]{Antoni Ramos-Buades}
\email{antoni.ramos-buades@uib.es}
\affiliation{\UIB}

\author[0000-0001-8798-0627]{Maximilian Dax}
\email{maximilian.dax@tuebingen.mpg.de}
\affiliation{\ELLIS}
\affiliation{\MPIIS}
\affiliation{\TUEAI}

\author[0000-0002-6987-6313]{Stephen R. Green}
\email{stephen.green2@nottingham.ac.uk}
\affiliation{\UoN}

\author[0009-0008-5938-6215]{Annalena Kofler}
\email{annalena.kofler@tuebingen.mpg.de}
\affiliation{\AEI}
\affiliation{\MPIIS}

\author[0000-0001-5154-8912]{Jakob Macke}
\email{Jakob.Macke@uni-tuebingen.de}
\affiliation{\MPIIS}
\affiliation{\MLS}

\author[0000-0002-8177-0925]{Bernhard Sch\"olkopf}
\email{bs@tuebingen.mpg.de}
\affiliation{\MPIIS}
\affiliation{\ELLIS}

\begin{abstract}

Multiple formation pathways have been proposed for the origin of binary black 
holes (BBHs). These include isolated binary evolution and dynamical assembly in
dense stellar environments such as nuclear or globular star clusters. Yet, the
fraction of BBHs originating from each channel remains uncertain. One way to
constrain this fraction is by investigating the orbital eccentricities of the BH
coalescences detected by the LIGO–Virgo–KAGRA (LVK) Collaboration. We analyze 84
BBHs from the first part of the fourth LVK observing run (O4a) using a
multipolar, eccentric, aligned-spin effective-one-body waveform model.  We
perform parameter inference with neural posterior estimation and nested
sampling. After incorporating astrophysical prior odds and comparing to the
quasicircular precessing-spin hypothesis, we find that no candidates reach a
high enough significance to claim a confident detection of eccentricity. We use
these upper limits to explore a model, in which all O4a BBHs originate from
single-single gravitational wave (GW) captures.  We perform hierarchical
inference on the velocity dispersion of the host environment of the BBHs and find $\sigma <
19.7\,\mathrm{km/s}$ (95\% credible upper bound). This disfavors single-single
capture in nuclear star clusters ($\sim 20-200\,\mathrm{km/s}$) as the dominant
source of all observed BBH mergers. Our analysis also jointly infers the mass, spin 
and redshift distributions and takes into account selection effects due to using 
quasi-circular templates for detection.
We verify that this dispersion bound does
not increase by repeating the inference on a synthetic catalog augmented with
eccentric events motivated by analyses of the third observing run of the LVK
(O3). Our results place improved constraints on the number of eccentric BBHs
and highlight the importance of eccentricity measurements in disentangling
compact-binary formation channels in current and future GW detectors.  
\end{abstract}



\section{Introduction}
\label{sec:introduction}
 \setcounter{footnote}{0} 

Through the first four LIGO-Virgo-KAGRA (LVK) observing runs (O1--O4a),
there have been 218 detections of compact binary coalescences
\citep{LIGOScientific:2018mvr, LIGOScientific:2020ibl, LIGOScientific:2019lzm, KAGRA:2021vkt, KAGRA:2023pio, LIGOScientific:2025slb, LIGOScientific:2025odr}. Despite the wealth of data, several formation channels
that could explain these mergers remain viable. A subset of these channels include: 

\begin{enumerate}
    \item Isolated binary evolution: this is when orbital separation between massive stars is reduced
    via dynamical friction from the common envelope \citep{1976IAUS...73...75P, 1976ApJ...208L..61O, Bethe:1998bn, Belczynski:2001uc, Dominik:2012kk},
    angular momentum transport via stable mass transfer \citep{1987ApJ...318..794H, 1997A&A...327..620S, King:1999me} or 
	because the stars remain compact and do not merge prematurely due to chemically homogeneous evolution \citep{Mandel:2015qlu, deMink:2016vkw}.  

	\item Field triples: here a system of three massive stars in the galactic field evolve according to
	stellar evolution and von~Zeipel-Kozai-Lidov (ZKL) oscillations 
    \citep{1910AN....183..345V, kozai1962secular, lidov1962evolution, giacaglia1964notes, Wen:2002km, Kimpson:2016dgk, Silsbee:2016djf, Antonini:2017ash, Fragione:2020nib, Arca-Sedda:2018qgq, Vigna-Gomez:2020fvw, Stegmann:2021jen}.
	These ZKL oscillations cause the inner binary to merge within a Hubble time. 

    \item Dynamical interactions in dense stellar clusters: here multi-body gravitational interactions of
    stars or compact objects decrease the orbital
    separation between the bodies
    \citep{heggie1975binary, Kulkarni:1993fr, sigurdsson1993primordial, Sigurdsson:1993tui, Sigurdsson:1994ju, PortegiesZwart:1999nm, Miller:2001ez, Gultekin:2005fd}.
    These can be further decomposed into
    binary-single or binary-binary interactions
    \citep{Samsing:2013kua, Samsing:2017rat, Zevin:2018kzq},
    single-single gravitational-wave captures
    \citep{OLeary:2005vqo, OLeary:2008myb, Gondan:2017wzd},
    and ZKL oscillations.

    \item Active galactic nuclei (AGN) disks: in this channel hardening of
    compact objects occurs within accretion disks around supermassive black
    holes \citep{2012MNRAS.425..460M, McKernan:2014oxa, Bartos:2016dgn, Stone:2016wzz, 2018ApJ...866...66M, 2019PhRvL.123r1101Y, 2020ApJ...898...25T, McKernan:2020lgr, Ford:2021kcw}.
    
    \end{enumerate}

To an extent, electromagnetic (EM) observations can be used to constrain the
fraction of binary black holes (BBHs) arising from these different formation
channels. For example, X-ray binaries provide an indirect measure of isolated
binary evolution \citep{Tauris:2003pf}.  A more direct EM probe of the common
envelope also exists in the form of Luminous red novae (LRNe)
\citep{2013A&ARv..21...59I,Matsumoto:2022qri}. However, these events are rare
\citep{Mason:2010tn, Tylenda:2010sh} and thus the rates of these transients are
low $7.8_{-3.7}^{+6.5} \: \text{Mpc}^{-3} \text{yr}^{-1}$
\citep{Karambelkar:2022tkx}. Importantly, both of these probes are restricted to
low redshifts ($z \lesssim 0.1$) \citep{Liotine:2022vwq}. They also 
cannot directly probe dynamically formed BBHs which do not necessarily emit EM
signatures during their formation. Furthermore, these X-ray binaries rarely lead to
BBH mergers \citep{Gallegos-Garcia:2022rve, Liotine:2022vwq, Romero-Shaw:2023mse}.

This can be remedied by using gravitational wave (GW) data from the
LVK network to determine which formation channels create BBH
mergers \citep{TheLIGOScientific:2014jea, VIRGO:2014yos, LIGOScientific:2016aoc, LIGOScientific:2018mvr, LIGOScientific:2020ibl, KAGRA:2020tym, LIGOScientific:2021usb, Capote:2024rmo, LIGO:2024kkz, LIGOScientific:2025slb, LIGOScientific:2025pvj}.  GWs interact weakly with
matter, and therefore act as a clean probe of the source properties of the BBH. 

While the aforementioned LVK collaboration studies have used the masses and spins of observed BBHs to constrain
formation channels, they have not yet taken into account orbital eccentricity (with the exception of \citealt{LIGOScientific:2025brd}).
Orbital eccentricity is important because it is a smoking-gun signature of
dynamical formation. Isolated binaries radiate away their eccentricity so
efficiently that the residual value upon entering the LVK detector band is only $\sim 10^{-11}$
\citep{peters1964gravitational}. In contrast, the eccentricity of dynamically formed
BBHs can remain significant ($\gtrsim 10^{-4}$) \citep{kozai1962secular, lidov1962evolution, heggie1975binary, Miller:2001ez, Wen:2002km, Samsing:2013kua}. Among dynamical channels, single-single GW captures are both
analytically tractable and produce the highest eccentricities, making them
the natural first channel to constrain using upper bounds on eccentricity from
individual events.

In this paper, we present the analysis of the 84 confidently detected BBH
signals from O4a using an eccentric waveform model. In Section \ref{sec:past_work}, we 
recap related work. In Section \ref{sec:methods},
we summarize the waveform models and parameter estimation techniques used. In
Section \ref{sec:results}, we present the posterior distributions and Bayes
factors for 84 events analyzed in O4a. We conclude that while there are several events which
have positive Bayes factors for eccentricity, no event reaches a high enough
significance level to claim a robust discovery. In this section we also show the
importance of glitch mitigation of these events and highlight the effect on the
event GW190701 from the third observing run (O3) of the LVK collaboration. In
Section \ref{sec:eccentricity_distribution} we describe the distribution of
eccentricities from single-single GW captures. In Section
\ref{sec:selection_function} we outline a method for including selection effects semi-analytically when the 
signal model differs from the template bank model. In Section \ref{sec:astro_implications}, we use the results from 
Sections \ref{sec:results}, \ref{sec:eccentricity_distribution} and \ref{sec:selection_function} to
perform a full population inference, jointly inferring the mass, spin, redshift, and
eccentricity distributions of the population. We show that combining eccentricity
constraints across events disfavors nuclear star clusters
as the dominant source of single-single captures. In Section
\ref{sec:discussion} we summarize our findings and discuss future work.


\section{Related work}
\label{sec:past_work}
There have been several inference studies for eccentric BBHs in the LVK data
\citep{Sun:2015bva, Romero-Shaw:2019itr, Romero_Shaw:2020thy, Wu:2020zwr, Lenon:2020oza, Ramos-Buades:2019uvh, Gayathri:2020coq, Romero-Shaw:2022xko, OShea:2021faf, Gamba:2021gap, Ramos-Buades:2023yhy, Bonino:2022hkj, Iglesias:2022xfc, Fei:2024ruj, Gupte:2024jfe, McMillin:2025hof, Planas:2025jny, Phukon:2025cky, Morras:2025xfu}.  Several events show positive evidence for eccentricity; however, when accounting for non-Gaussian
noise or spin-precession, no \textit{individual event} reaches statistical
significance for a confident detection.

These studies usually compute the
Bayes factor between a quasicircular precessing-spin (QCP) model and an eccentric aligned-spin (EAS) model (hereafter
$\mathcal{B}_{\text{EAS/QCP}}$).
Following \cite{Gupte:2024jfe}, while the BBH event GW200129 supports the eccentric hypothesis, its $\log_{10} \mathcal{B}_{\text{EAS/QCP}}$ 
ranges between 0.76 and 4.92 depending on the subtraction method used for non-Gaussian 
noise ``glitches'' in the detector. The BBH event GW190701 
also supports the eccentric hypothesis, with a $\log_{10} \mathcal{B}_{\text{EAS/QCP}}$ of 3.0, but 
there have not been studies investigating the properties of the non-Gaussian noise 
of the event. The BBH event GW200208\_22 also shows
signatures of eccentricity, but it has a high False Alarm Rate (FAR) and its $\log_{10} \mathcal{B}_{\text{EAS/QCP}}$ 
is below 2 (\cite{Romero-Shaw:2022xko, McMillin:2025hof, Romero-Shaw:2025vbc}). There is also the neutron star black hole (NSBH) event, GW200105, which has some support for eccentricity (\cite{Fei:2024ruj, Morras:2025xfu}). However, this event also reaches only modest Bayes factors of $\log_{10} \mathcal{B}_{\text{EAS/QCP}} = 1.2$ \citep{Kacanja:2025kpr, Planas:2025plq}. GW190521 has also shown
signs of eccentricity in \cite{Romero_Shaw:2020thy, Gayathri:2020coq, Gamba:2021gap}, however, more recent waveform models and parameter estimation methods do not recover these features
\citep{Ramos-Buades:2023yhy, Iglesias:2022xfc, Gupte:2024jfe, Gamboa:2024hli}
\footnote{We note, though, these studies do not probe eccentricity in the merger, and therefore are not immediately in contention 
with \cite{Gayathri:2020coq}. }. There are also other GW events such as GW190620, GW190929
and GW191109 which have mild support for eccentricity with $\log_{10} \mathcal{B}_{\text{EAS/QCP}}$ between 0.1 and 
1. Finally, while high likelihood eccentric modes exist in all of these events, the choice of prior can often drastically reduce the support for eccentricity.
This problem was partially addressed in \cite{Gupte:2024jfe} where
inference was performed on the rate of eccentric and quasicircular BBHs
to compute the probabilities that individual events were eccentric and to 
compute the probability of eccentricity in the population. 

Recently, \cite{Xu:2025xxx, Malagon:2026uev} and \cite{Zeeshan:2026pga} have performed
concurrent and independent analyses of O4a with eccentric waveform models, with the
latter also probing the population distribution. We discuss comparisons with their results and ours 
in section \ref{sec:discussion}.


\section{Methods}
\label{sec:methods}

We start by summarizing the basics of the Bayesian inference formalism used in GW astronomy. 
Our hypothesis, $\mathcal{H}_a$, is that in the detector data, $d$, an observed GW signal is
described by a waveform model, $a$, with parameters $\bm{\vartheta}_a$. 
The posterior probability distribution on the GW parameters is obtained using 
Bayes' theorem
\begin{equation}\label{eq:Bayes_theorem}
p(\bm{\vartheta}_a|d, \mathcal{H}_a) = \frac{p(d|\bm{\vartheta}_a, \mathcal{H}_a)\,p(\bm{\vartheta}_a|\mathcal{H}_a) }{p(d|\mathcal{H}_a)} \,.
\end{equation}
Here, $p(\bm{\vartheta}_a|\mathcal{H}_a)$ is the prior probability distribution,
$p(d|\bm{\vartheta}_a, \mathcal{H}_a)$ is the likelihood, and
$p(d|\mathcal{H}_a)$ is the evidence of the hypothesis $\mathcal{H}_a$.
For the analysis of stationary Gaussian noise, the likelihood takes the form of the Whittle likelihood \citep{Whittle:1951ppp}. One can
compute Bayes factors by taking the ratio of model evidences, $\mathcal{B}_{a/b}
= p(d|\mathcal{H}_a) / p(d|\mathcal{H}_b)$. In this paper we consider the EAS,
quasicircular aligned-spin (QCAS) and QCP hypotheses. We summarize the priors used in this work 
in Table \ref{tab:priors}.

A central concern is the choice of prior distribution for the eccentricity parameter. Based 
on isolated binary or dense stellar cluster simulations, we expect several Gaussian-like features in
the regime $\log_{10} e_{\text{10Hz}} = -11$ to $\log_{10} e_{\text{10Hz}} = 0$,
corresponding to different dynamical and isolated formation channels \citep{Belczynski:2001uc, Kremer:2019iul, Olejak:2022zee}. One could conservatively smooth these into a uniform prior
on $\log_{10} e_{\text{10Hz}}$ setting the lower bound to $\log_{10} e_{\text{10Hz}} =
-11$. However, the posterior will then strongly depend on the choice of the lower
bound of the prior. 
We therefore use a uniform prior in
$e_{\text{10Hz}}$. This avoids the strong sensitivity to the lower bound
that arises with a log-uniform prior, which can dominate the evidence integral
and the resulting Bayes factors \citep{Gupte:2024jfe}. However, it implicitly assumes higher 
eccentricities than expected from astrophysics.

To recover the astrophysical knowledge at the individual-event level, we
compute odds ratios $\mathcal{O}_{\text{EAS/QCAS}}$ by
multiplying the Bayes factor $\mathcal{B}_{\text{a/b}}$ with the prior odds,
$p(H_\text{a}) / p(H_\text{b})$ (see \cite{Gupte:2024jfe} for a more rigorous
treatment). Here, the prior odds are simply the relative rate of ``measurably
eccentric'' (i.e., $e_{\text{10Hz}} > 0.05$ \citep{Lower:2018seu}) and quasicircular channels, $R_{\text{EAS}} /
R_{\text{QCAS}}$\footnote{Note this $e_{\text{10Hz}} > 0.05$ threshold is for a GW150914-like injection. A general treatment, 
would set a mass and signal-to-noise dependent threshold.}.  We use the relative rates compiled from the astrophysics
literature which indicate the prior odds are $R_{\text{EAS}} / R_{\text{QCAS}}
= 0.006^{+0.023}_{-0.005}$ (median and 95\% credible interval of the induced
prior on the ratio)\footnote{A previously quoted value of $R_{\text{EAS}} / R_{\text{QCAS}} \approx 0.023$
corresponds to the ratio evaluated at the joint maximum probability of $p(R_{\text{EAS}},
R_{\text{QCAS}})$. Here we take the credible interval of the induced prior instead since 
we are not marginalizing over the rates. } \citep{Gupte:2024jfe}, i.e.\ between $\sim 0.1\%$ and $\sim 3\%$ of
BBHs are measurably eccentric (95\% credible interval).
Combining the uniform eccentricity prior with these prior odds is approximately equivalent to
adopting a population-level prior that places a fraction $\alpha$ of the weight
below the threshold $e_{\text{10Hz}} = 0.05$ and $1 - \alpha$ above
it.\footnote{This equivalence is approximate because the EAS model has support
over all of $e_{\text{10Hz}} \in [0, 0.8]$, not only above the threshold. However,
waveforms with $e_{\text{10Hz}} < 0.05$ are nearly indistinguishable from
quasicircular ones at current detector sensitivities, so the Bayes factor is
effectively driven by the $e_{\text{10Hz}} > 0.05$ region.}
In our analysis $\alpha$ is set by the astrophysical rate estimates, but
different readers may prefer different values based on their own expectations;
by reporting Bayes factors computed with a uniform prior, we provide a baseline
that can be combined with any choice of prior odds.
In Section~\ref{sec:astro_implications}, we go further and replace the
uniform prior with a physically motivated forward model, reinterpreting the
individual-event posteriors in a population-level context.

\begin{table}[t]
\renewcommand{\arraystretch}{1.3}
\setlength{\tabcolsep}{4pt}
\begin{center}
\begin{tabular}{lll}
\toprule
Parameter & Prior & Range \\
\midrule
\multicolumn{3}{l}{\textit{Intrinsic parameters}} \\
$\mathcal{M}$ & Uniform in comp.\ masses & $[7, 200]~\text{M}_\odot$ \\
$q$ & Uniform in comp.\ masses & $[0.125, 1]$ \\
$\chi_1$ & Uniform & $[-0.99, 0.99]$ \\
$\chi_2$ & Uniform & $[-0.99, 0.99]$ \\
$e_{10\text{Hz}}$ & Uniform & $[0, 0.8]$ \\
$\log_{10} e_{10\text{Hz}}$ & Uniform$^\dagger$ & $[-4, -0.1]$ \\
$\zeta_{10\text{Hz}}$ & Uniform (periodic) & $[0, 2\pi]$ \\
\midrule
\multicolumn{3}{l}{\textit{Extrinsic parameters}} \\
$\phi_c$ & Uniform & $[0, 2\pi]$ \\
$\theta_{JN}$ & Sine & $[0, \pi]$ \\
$d_L$ & UniformComovingVolume & $[100, 15000]~\text{Mpc}$ \\
$\alpha$ & Uniform & $[0, 2\pi]$ \\
$\delta$ & Cosine & $[-\pi/2, \pi/2]$ \\
$\psi$ & Uniform & $[0, \pi]$ \\
$t_c$ & Uniform & $[-0.1, 0.1]~\text{s}$ \\
\bottomrule
\end{tabular}
\end{center}
\caption{\label{tab:priors} Summary of prior distributions used in the parameter estimation.
The parameters are: detector frame chirp mass $\mathcal{M}$, mass ratio $q = m_2/m_1 \leq 1$,
dimensionless aligned-spin components $\chi_1$ and $\chi_2$,
orbital eccentricity $e_{10\text{Hz}}$, relativistic anomaly $\zeta_{10\text{Hz}}$,
phase at coalescence $\phi_c$, inclination angle $\theta_{JN}$,
luminosity distance $d_L$, right ascension $\alpha$, declination $\delta$,
polarization angle $\psi$, and geocenter time $t_c$.
The chirp mass and mass ratio are sampled with a prior that is uniform in the component masses.
The component mass constraints depend on the \texttt{DINGO} network: $m_{1,2} \in [10, 120]~\text{M}_\odot$
for the high-mass network and $m_{1,2} \in [1, 120]~\text{M}_\odot$ for the low-mass network.
Eccentricity and relativistic anomaly are defined at an orbit-averaged detector-frame reference frequency of 10~Hz.
$^\dagger$The log-uniform prior on eccentricity is used for the population-level inference in Section~\ref{sec:astro_implications}.}
\end{table}

In this work, we employ three different waveform approximants from the \texttt{SEOBNRv5} family of models \citep{Pompili:2023tna, Ramos-Buades:2023ehm, Khalil:2023kep, vandeMeent:2023ols, Mihaylovv5, Gamboa:2024imd, Gamboa:2024hli, Haberland:2025luz, Estelles:2025zah}.
Namely, the QCAS model
\texttt{SEOBNRv5HM} \citep{Pompili:2023tna}, the QCP model \texttt{SEOBNRv5PHM} \citep{Ramos-Buades:2023ehm}, and the EAS
model \texttt{SEOBNRv5EHM} \citep{Gamboa:2024imd, Gamboa:2024hli}.
The eccentricity study in \cite{Gupte:2024jfe} was performed with the previous-generation \texttt{SEOBNRv4EHM} model \citep{Khalil:2021txt,Ramos-Buades:2021adz}.
Here, we use the \texttt{SEOBNRv5EHM} model due to its improved accuracy, faster computational performance, and robust waveform generation.
Particularly, it was shown in \cite{Gamboa:2024hli} that \texttt{SEOBNRv5EHM} is, overall, one-order-of-magnitude more accurate than \texttt{SEOBNRv4EHM} and other state-of-the-art EAS waveform models \citep{Nagar:2024dzj,Planas:2025feq}.
The main differences of \texttt{SEOBNRv5EHM} with respect to \texttt{SEOBNRv4EHM} are threefold.
First, the underlying quasicircular model baselines are improved, including a refined effective-one-body (EOB) Hamiltonian and better calibration to quasicircular numerical-relativity waveforms compared to \texttt{SEOBNRv4HM} \citep{Cotesta:2018fcv}.
Second, nonspinning eccentricity corrections are included in the radiation-reaction force (absent in \texttt{SEOBNRv4EHM}) and in the waveform modes, now up to third post-Newtonian (PN) order instead of second.
Third, the model is coupled with the full 3PN evolution equations for the parameters $(x, e, \zeta)$.
These parameters come from a Keplerian parametrization of the orbit \citep{darwin1959gravity},
\begin{equation}
\label{eq:qc_parameterization}
r = \frac{p}{1 + e \cos \zeta},
\end{equation}
where $r$ is the relative separation of the binary, $p$ is the semi-latus rectum, $e$ is the Keplerian eccentricity, and $\zeta$ is the relativistic anomaly. The semi-latus rectum can be related to the dimensionless orbit-averaged orbital frequency $x = \langle M \Omega \rangle^{2/3}$, where $\Omega$ is the instantaneous angular frequency of the binary.
Since these orbital parameters evolve with the binary, we choose to report them at an orbit-averaged detector-frame reference frequency of 10 Hz;
thus, we denote them as $e_{10\text{Hz}}$ and $\zeta_{10\text{Hz}}$.
Furthermore, since the orbital eccentricity is subject to the choice of coordinates in general relativity, for events with high eccentricity, we extract the gauge-invariant eccentricity from the waveform using the \texttt{gw\_eccentricity} package \citep{Shaikh:2023ypz} and label this $e_{\text{gw}}$. We report $e_{\text{gw}}$ at the dimensionless frequency $M_{\text{tot}} f_{\text{ref}}$, where $M_{\text{tot}}$ is the total detector frame mass of the binary and $f_{\text{ref}}$ is the reference frequency of 10 Hz. By multiplying by the total mass of the binary we ensure our measurement point of the eccentricity is redshift independent.

The improvements of \texttt{SEOBNRv5EHM} with respect to \texttt{SEOBNRv4EHM}
allowed us to extend the prior on the eccentricity to $e_{10\text{Hz}} \in [0,
0.8]$ from the $e_{10\text{Hz}} \in [0, 0.5]$ used in \cite{Gupte:2024jfe}. 
Note, however, that this is not a strict increase to $e_{10\text{Hz}} = 0.8$
across the entire parameter space.  Highly eccentric binaries are challenging to
model because of the strong-field, high-velocity configurations reached during
close encounters.  Thus, initializing a binary with a small relative
separation or very close to merger can lead to inaccuracies in waveform generation due to the breakdown of the underlying model's approximations. This is particularly problematic at periastron ($\zeta = 0$) for large $e$ and for
high-total-mass systems, where the analysis focuses only on the final portion of
the waveform observable by LVK detectors. 
The \texttt{SEOBNRv5EHM} model ameliorates this problem by performing a backward integration of a set of secular evolution equations in case the requested initial state of the binary corresponds to a strong-field configuration, as explained in \cite{Gamboa:2024imd}.
However, even with this method, there still remain extreme binary configurations which cannot be generated.
Therefore, the
effective prior on $e_{10\text{Hz}}$ is conditional on the total mass and
relativistic anomaly.  To assess posterior railing, we construct a concave
hull of the parameter samples that have a successful waveform generation using \texttt{alphashape} \citep{bellock:2021}. We then plot this concave
hull on the $(e_{10\text{Hz}}, \zeta)$ 2D
marginal for each posterior given the total mass of the event. If the posterior
approaches this region, this would indicate railing against the model's boundary
of successful waveform generations (see the orange bands in
Fig.~\ref{fig:posteriors}).

We obtain posterior distributions using the 
inference codes \texttt{DINGO} \citep{Dax:2021tsq} and
\texttt{Bilby} \citep{Ashton:2018jfp, Romero-Shaw:2020owr}.
Traditional stochastic sampling methods such as nested sampling, as
implemented in \texttt{Bilby}, require $\mathcal{O}(\text{days--weeks})$
per event, making a full catalog analysis computationally prohibitive.
We therefore use \texttt{DINGO} for the majority of events.
Once trained, \texttt{DINGO} uses an embedding network,
normalizing flow and symmetries of the GW~parameters to deliver proposal
posterior distributions in $\mathcal{O}$(minutes). These proposal posteriors are
then importance sampled in $\mathcal{O}$(hours) to ensure agreement with the
true posterior distribution under the Gaussian noise assumption and to obtain
Bayesian evidences \citep{Dax:2022pxd}. Importance sampling comes with a natural
heuristic to assess sampling convergence, the \textit{effective sample size}, 
which we denote $n_{\text{eff}}$.  For all events in this paper, we aim to obtain
a minimum of 3,000 effective samples. If this criterion cannot be met
(usually due to glitches or low masses, see~\cite{Dax:2022pxd}), we use nested
sampling via the \texttt{dynesty} sampler~\citep{Speagle:2019ivv} accessed
through \texttt{Bilby} \citep{Ashton:2018jfp}. We use 1024 live points for the \texttt{dynesty} runs.
Additionally, for events exhibiting posterior railing or changes in the detector
frequency range, we analyze these events with \texttt{Bilby} to reduce the
number of trained networks required (note that the problem of changing frequency ranges was
recently addressed in~\cite{Kofler:2025dux}). Events with a chirp mass between 6
$M_\odot$ and 15 $M_\odot$ are analyzed using the multibanding
technique~\citep{Vinciguerra:2017,Morisaki:2021}, which was introduced to
\texttt{DINGO} in \cite{dax2024bns, Kofler:2025dux}.

To mitigate the impacts of glitches, we marginalize over inferred glitch
realizations for selected events. This requires us to first jointly infer the signal and glitch
properties using a quasicircular waveform model for the BBH \citep{Cornish:2014kda, Udall:2022vkv}. The glitch is modeled as either parameterized
fast scattering (for GW190701) or a sum of wavelets (for GW231114\_043211,
GW231221\_135041, and GW231223\_032836). In the case 
of the sum-of-wavelets model, \texttt{BayesWave} internally uses trans-dimensional Monte Carlo Markov Chain (MCMC). From these
inferred glitch posteriors, we draw 100 realizations and subtract them from the
strain data, perform parameter estimation on each, and combine the
posteriors inferred accordingly. 

\section{Eccentric analyses of O4a BBH signals}
\label{sec:results}

\begin{figure*}
\includegraphics[width=\textwidth]{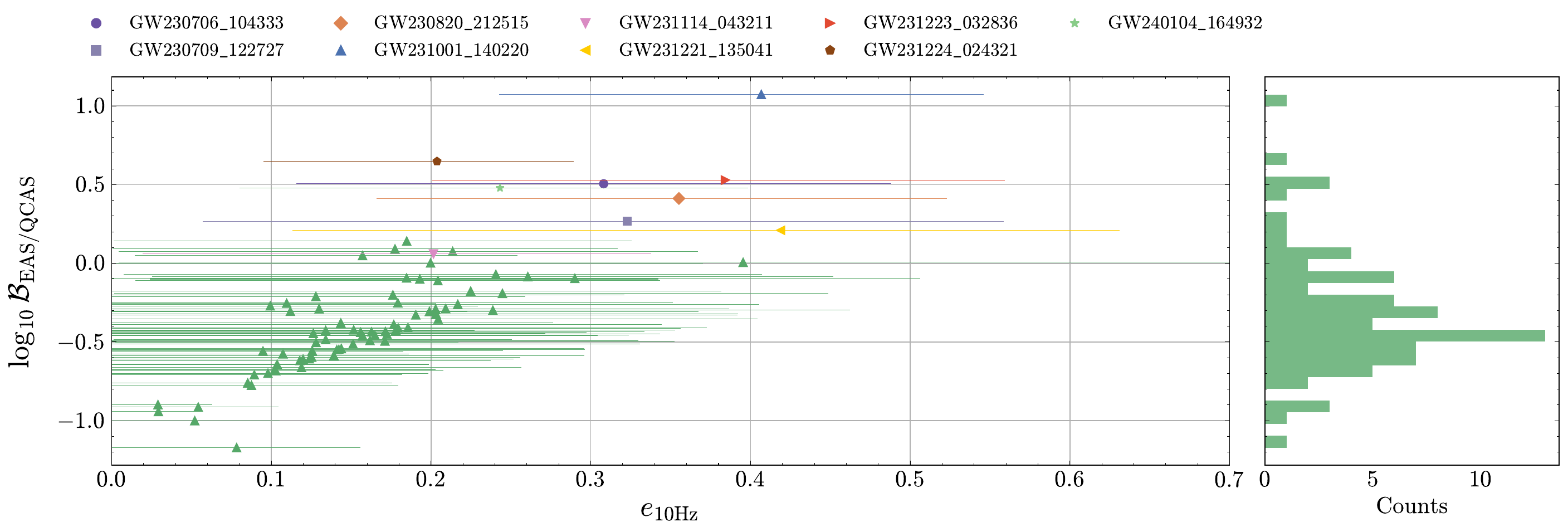}
\caption{\label{fig:bf_lineplot} \textit{Left} $\log_{10} \mathcal{B}_{\text{EAS/QCAS}}$ and the $e_{\text{10Hz}}$ 90\% highest-density intervals (HDIs)
for the 84 events analyzed from O4a. \textit{Right} is a histogram counting the distribution 
of $\log_{10} \mathcal{B}_{\text{EAS/QCAS}}$. A few of these events exclude $e_{\text{10 Hz}} = 0$
at the 90\% level and have a $\log_{10} \mathcal{B}_{\text{EAS/QCAS}} > 0.2$. We also highlight 
GW231114\_043211 which has a significant reduction in eccentricity when marginalizing over the glitch. 
We use different 
colors for these events to guide the eye. Despite a few positive Bayes factors, when taking into account the prior 
odds, none of these events can be considered confidently eccentric. }

\end{figure*}

In Fig.~\ref{fig:bf_lineplot} we present the analysis of the 84 GW events from
O4a\footnote{Posterior samples, trained networks, hierarchical inference results and selection function injection sets are available at \url{https://doi.org/10.5281/zenodo.21221337}. Code to reproduce all figures and tables is available at \url{https://github.com/nihargupte-ph/o4a-eccentricity}.} \footnote{We do not analyze GW230630\_070659, because it is likely of
instrumental origin, nor the NSBH GW230518\_125908, which occurred during the
engineering run \citep{LIGOScientific:2025slb}.}. We marginalize over the calibration 
envelope, use segment lengths of 16 seconds and use Welch PSDs. Of the 84 events, we further 
investigate the 9 with the highest Bayes factors ($\log_{10}
\mathcal{B}_{\text{EAS/QCAS}} > 0.2$ before glitch marginalization). None of these events has a $\log_{10}
\mathcal{O}_{\text{EAS/QCAS}}$ exceeding 0, and therefore none has
sufficient support to claim a detection of eccentricity. 
We summarize the source properties and Bayes factors of these 9~events in
Table~\ref{tab:interesting_table} and show their eccentricity vs anomaly posterior distributions in the \hyperref[sec:appendix]{appendix}. We report the eccentricities of all 84~events analyzed in Table~\ref{tab:all_events} in the \hyperref[sec:appendix]{appendix}.


\begin{table*}[t]
\begin{center}
\setlength{\tabcolsep}{4pt}
\renewcommand{\arraystretch}{1.5}
\begin{tabular}{lCCCCCCCCC}
	\toprule
	Event name & $\text{FAR}_{\text{min}}$ ($\text{yr}^{-1}$) & $\log_{10} \mathcal{B}_{\text{EAS/QCAS}}$ & $\log_{10} \mathcal{B}_{\text{EAS/QCP}}$ & $\log_{10} \mathcal{O}_{\text{EAS/QCAS}}$ & $e_{10\text{Hz}}$ & $e_{\text{gw, 10Hz}}$ & $\chi_{\text{eff}}$ & $\text{M}_{\text{src, total}} [ M_\odot ]$ \\
	\midrule
		GW230706\_104333 & $0.2$ & \ColorFour$0.51$ & \ColorFour$0.40$ & \ColorOne$-1.7^{+0.7}_{-0.7}$ & $0.31_{-0.19}^{+0.18}$ & $0.30_{-0.17}^{+0.19}$ & $0.11_{-0.15}^{+0.15}$ & $27.3_{-3.3}^{+3.3}$ \\
		GW230709\_122727 & $7.1 \times 10^{-2}$ & \ColorFour$0.27$ & \ColorThree$0.19$ & \ColorOne$-2.0^{+0.7}_{-0.7}$ & $0.32_{-0.27}^{+0.24}$ & $0.33_{-0.28}^{+0.23}$ & $0.05_{-0.31}^{+0.31}$ & $72.3_{-15.2}^{+15.7}$ \\
		GW230820\_212515 & $0.2$ & \ColorFour$0.41$ & \ColorFour$0.50$ & \ColorOne$-1.8^{+0.7}_{-0.7}$ & $0.36_{-0.19}^{+0.17}$ & $0.34_{-0.17}^{+0.16}$ & $0.10_{-0.30}^{+0.30}$ & $98.0_{-15.4}^{+16.1}$ \\
		GW231001\_140220 & $1.6 \times 10^{-5}$ & \ColorFive$1.07$ & \ColorFour$0.62$ & \ColorOne$-1.2^{+0.7}_{-0.7}$ & $0.41_{-0.16}^{+0.14}$ & $0.34_{-0.13}^{+0.12}$ & $0.05_{-0.29}^{+0.38}$ & $107.3_{-21.7}^{+23.5}$ \\
		GW231114\_043211 & $1.3 \times 10^{-4}$ & \ColorThree$0.06$ & \ColorFour$0.56$ & \ColorOne$-2.2^{+0.7}_{-0.7}$ & $0.20_{-0.18}^{+0.14}$ & $0.20_{-0.18}^{+0.14}$ & $0.19_{-0.14}^{+0.15}$ & $36.1_{-5.4}^{+5.3}$ \\
		GW231221\_135041 & $0.5$ & \ColorThree$0.21$ & \ColorThree$-0.07$ & \ColorOne$-2.0^{+0.7}_{-0.7}$ & $0.42_{-0.31}^{+0.21}$ & $0.38_{-0.27}^{+0.23}$ & $0.21_{-0.34}^{+0.34}$ & $74.4_{-17.0}^{+17.1}$ \\
		GW231223\_032836 & $3.8 \times 10^{-4}$ & \ColorFour$0.53$ & \ColorFour$0.50$ & \ColorOne$-1.7^{+0.7}_{-0.7}$ & $0.38_{-0.18}^{+0.17}$ & $0.37_{-0.20}^{+0.19}$ & $0.02_{-0.38}^{+0.36}$ & $77.1_{-15.8}^{+14.6}$ \\
		GW231224\_024321 & $< 1.0 \times 10^{-5}$ & \ColorFour$0.65$ & \ColorFour$0.49$ & \ColorOne$-1.6^{+0.7}_{-0.7}$ & $0.20_{-0.11}^{+0.08}$ & $0.20_{-0.11}^{+0.09}$ & $-0.04_{-0.07}^{+0.06}$ & $16.6_{-0.9}^{+0.9}$ \\
		GW240104\_164932 & $< 1.0 \times 10^{-5}$ & \ColorFour$0.48$ & \ColorThree$-0.13$ & \ColorOne$-1.7^{+0.7}_{-0.7}$ & $0.24_{-0.17}^{+0.15}$ & $0.24_{-0.16}^{+0.15}$ & $0.09_{-0.16}^{+0.17}$ & $72.6_{-8.7}^{+9.3}$ \\

	\bottomrule
\end{tabular}
\end{center}
\caption{\label{tab:interesting_table} Table of the 9 BBH signals in O4a which were selected based on having some support for
eccentricity ($\log_{10} \mathcal{B}_{\text{EAS/QCAS}} > 0.2$) before glitch
marginalization. Note this threshold was chosen just to highlight a few events and does not 
imply that events with $\mathcal{B}_{\text{EAS/QCAS}} > 0.2$ are eccentric. The reported values are obtained after glitch marginalization;
notably, GW231114\_043211 drops below the selection threshold after
marginalization (from $\log_{10} \mathcal{B}_{\text{EAS/QCAS}} = 0.36$ to
$0.06$) but is retained for completeness. We compute Bayes factors comparing the
EAS hypothesis to the QCAS and QCP hypotheses and report the results in columns
3 and 4. We also multiply by the prior odds which are
$R_{\text{EAS}}/R_{\text{QCAS}} = 0.006^{+0.023}_{-0.005}$, 95\% credible interval. This uncertainty
on the prior odds is induced by priors on the rates of mergers described in \cite{Gupte:2024jfe}.
We report the resulting odds ratios in column 5.
While these 9 events mildly support the eccentric hypothesis in the Bayes factor, the odds
ratios show that individually they are not confidently eccentric. In columns 6-9
we report the highest density credible intervals for the eccentricity, ``GW
eccentricity'', effective spin and source frame total mass.  } 
\end{table*}

It is important to consider the possibility of non-Gaussian noise (also known as
``glitches'')  biasing the parameter inference of eccentricity. Of the 9 events
presented in Table~\ref{tab:interesting_table}, 4 of them---GW230709\_122727,
GW231114\_043211, GW231221\_135041, GW231223\_032836---have associated glitches. 
Rather than subtracting a single realization of the glitch from the data, as is done in \cite{LIGOScientific:2025slb}, we
marginalize over the glitch posterior as described in Section~\ref{sec:methods}.

This is important because subtracting a single glitch realization
ignores the uncertainty in the glitch model, which can lead to large
biases in parameter estimation \citep{Payne:2022spz, Udall:2024ovp, Udall:2025bts}. One unresolved aspect
of glitch mitigation is the reliance on a quasicircular signal model during trans-dimensional MCMC.
If the true signal were eccentric, one could overestimate the glitch distribution and in turn 
bias the source parameter estimation. We leave this problem for future work.

\begin{figure*}
\includegraphics[width=\textwidth]{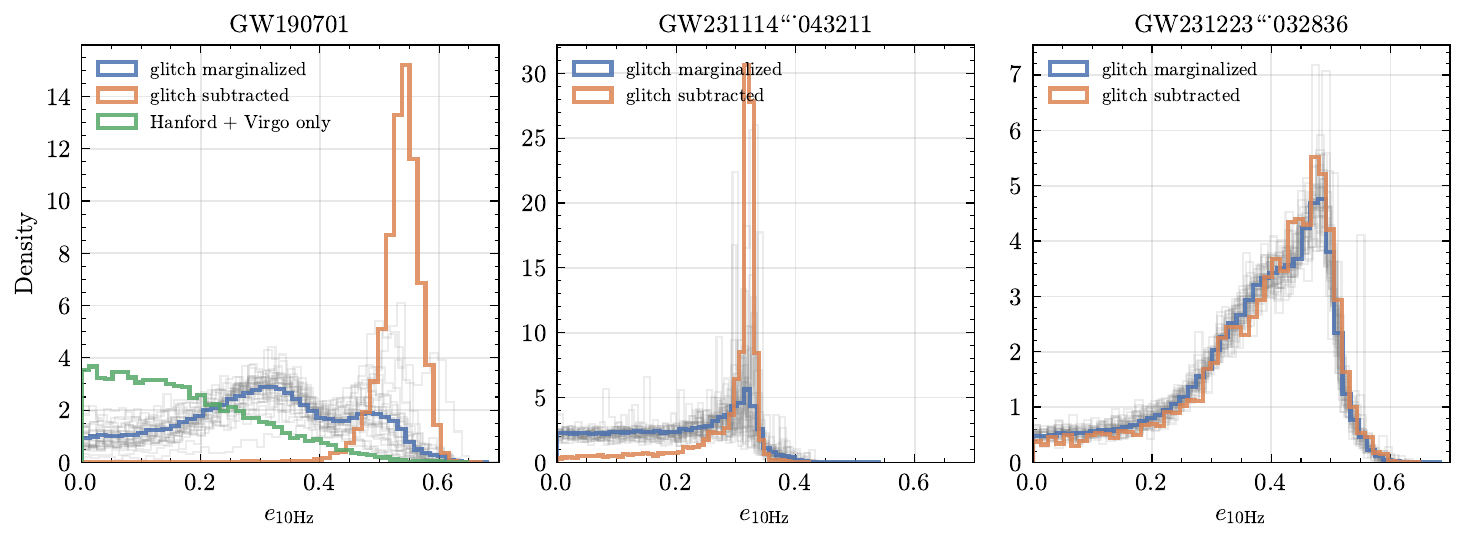}
\caption{\label{fig:glitch_marginalizations} Posterior distributions on $e_{10\text{Hz}}$ for GW190701, GW231114\_043211,
and GW231223\_032836 when either subtracting the glitch (orange) or marginalizing over 100 glitch realizations (blue).
For GW190701, when marginalizing over the glitch instead of taking one fair draw from the glitch distribution, two modes for eccentricity appear, but the distribution has support around zero $e_{10\text{Hz}} \sim 0$. When using the glitch marginalized posterior, 
the Bayes factors for GW190701 drop from $\log_{10}
\mathcal{B}_{\text{EAS/QCAS}} = 3.68$ to $\log_{10}
\mathcal{B}_{\text{EAS/QCAS}} = 0.42$. For GW231114\_043211, the Bayes
factors change from $\log_{10} \mathcal{B}_{\text{EAS/QCAS}} = 0.36$ to $\log_{10} \mathcal{B}_{\text{EAS/QCAS}} = 0.06$,
while the change for GW231223\_032836 is within the statistical error for the Bayes factor.
}

\end{figure*}

\citet{Udall:2025bts}
found that in the case of spins, the glitch affects parameter estimation when it overlaps with the 
signal track. Therefore, we expect biases from subtracting the 
glitch in GW231114\_043211 and GW231223\_032836, as well as the O3 events GW190701 and GW200129. 
Since the glitch subtraction of GW200129 was explored in \cite{Payne:2022spz,Gupte:2024jfe}, 
we focus here on GW231114\_043211, GW231223\_032836, and GW190701. We present the 
results of glitch marginalization for these events in Fig.~\ref{fig:glitch_marginalizations},
where GW190701 shows a large increase in support for the $e_{\text{10Hz}} \sim
0$ region after including the marginalization. We can compute Bayes factors using
the Savage-Dickey ratio and find $\log_{10} \mathcal{B}_{\text{EAS/QCAS}} =
0.42$ when using the glitch marginalized posterior. This is in contrast to using
the glitch subtracted posterior, where we find $\log_{10}
\mathcal{B}_{\text{EAS/QCAS}} = 3.68$. For GW231114\_043211, the Bayes factors change from $\log_{10} \mathcal{B}_{\text{EAS/QCAS}} = 0.36$ to $\log_{10} \mathcal{B}_{\text{EAS/QCAS}} = 0.06$. For GW231223\_032836, the effect is less pronounced with the change within the statistical error of the Bayes factor.

It remains to be seen whether GW190701 has high support for eccentricity.
If the true signal were eccentric, a quasicircular signal model in the
trans-dimensional MCMC would fail to capture the excess
signal-to-noise ratio (SNR) from eccentricity. This residual power could then be absorbed by the glitch model. However, the presented results
show that this event's Bayes factor is significantly lower when using glitch marginalization as opposed to
glitch subtraction. This is driven by the similarity between the scattered-light glitch and possible periastron passages in the Livingston data. This highlights the importance of glitch
marginalization in parameter estimation studies especially when eccentricity is
considered.

We note that a glitch with an SNR less than 5 is observed in Hanford for GW230709\_122727 which 
cannot be taken into account by trans-dimensional MCMC. Therefore, instead of using glitch marginalization, we increase the
$f_{\text{min}}$ in the likelihood integral in Hanford to 50~Hz. Note that this
removes the inspiral portion of the signal for Hanford meaning that only
Livingston can contribute to the measurement of eccentricity. Nonetheless, a
high likelihood mode at $e_{\text{10Hz}} \sim 0.34$ still exists in the posterior for this event
as shown in Table~\ref{tab:interesting_table} and Fig.~\ref{fig:posteriors}.

GW231221\_135041 has $e_{\text{10Hz}} = 0.42_{-0.31}^{+0.21}$ and is one of the
four GW events identified in O4a with a FAR $< 1\,\text{yr}^{-1}$ exclusively by
the cWB-BBH search pipeline \citep{Mishra:2022ott}. For these four events, the
coherent SNR reported by cWB-BBH is higher than that obtained with
matched-filtering analyses \citep{LIGOScientific:2025slb}, which may indicate
physics not included in the matched-filter templates, such as orbital
eccentricity. However, when considering the astrophysical odds, we find
$\log_{10} \mathcal{O}_{\text{EAS/QCAS}} = -2.0$ for GW231221\_135041,
indicating no significant support for eccentricity. There is also no support for
eccentricity for the other three cWB-BBH-only events (GW230712\_090405,
GW231004\_232346, GW231230\_170116), with $\log_{10}
\mathcal{B}_{\text{EAS/QCAS}}$ of $-0.55$, $-0.36$, and $-0.09$, respectively.

\section{The eccentricity distribution from single-single GW captures}
\label{sec:eccentricity_distribution}

We now describe a method to compute the distribution of eccentricity 
expected in the LVK band assuming formation from single-single GW captures. This model will be used in 
Section~\ref{sec:astro_implications} to draw astrophysical conclusions.

In the single-single GW capture scenario,
two black holes on initially unbound orbits pass close enough to lose sufficient
energy through GW emission to become bound
\citep{peters1964gravitational, 1989ApJ...343..725Q}. 
The energy radiated during a close encounter scales steeply with the component
masses, so more massive binaries radiate a greater fraction of their kinetic
energy at pericenter, allowing capture from wider orbits.
The resulting eccentricity at a given reference frequency depends on $\sigma$
as well as the total detector frame mass $M$ and detector frame reduced mass $\mu$ of the binary.

We compute the forward model $p(e | \sigma, M, \mu)$ using a Monte Carlo
procedure. The distribution can be written as
\begin{equation}
	\label{eq:velocity_slices}
	p(e | \sigma, M, \mu) = \int dv_\infty p(e | M, \mu, v_\infty) p(v_\infty | \sigma).
\end{equation}
Here $v_\infty$ is the relative speed of two BHs at infinity. Note all of these distributions 
are implicitly conditioned on the ``capture'' condition, we do not consider scattering orbits here and hence 
$e < 1$.
The distribution for $v_\infty$ is given
by a Maxwellian distribution with dispersion $\sqrt{6}\,\sigma$
(the $\sqrt{6}$ factor accounts for both the isotropic 3D and relative velocity factors),
weighted by the capture cross-section. This gives an
effective distribution
\begin{equation}
\label{eq:p_vinf}
p(v_\infty | \sigma) \propto v_\infty^{3/7} \exp\left(-\frac{v_\infty^2}{4\sigma^2}\right).
\end{equation}
The $v_\infty^{3/7}$ exponent arises from combining the Maxwell-Boltzmann
distribution ($\propto v_\infty^2$) with the rate of captures ($\propto
v_\infty^{-11/7}$). The rate of captures comes from the velocity dependence of the
gravitationally focused capture cross-section ($\propto v_\infty^{-18/7}$) and
the encounter flux ($\propto v_\infty$) \cite{1989ApJ...343..725Q}.

To compute $p(e | M, \mu, v_\infty)$, for each $v_\infty$, we first write it as an 
integral over the pericenter distance $r_p$,
\begin{equation}
	\int d r_p p(e | M, \mu, r_p) p(r_p | v_\infty).
\end{equation}
$p(r_p | v_\infty)$ is solely determined by the
maximum pericenter distance, $r_{p,\rm max}$, that allows capture. This in turn is
determined by requiring that the GW energy radiated during the close encounter
\citep{peters1964gravitational} exceeds the kinetic energy at infinity, $\Delta
E_{\rm GW} \geq \frac{1}{2}\mu v_\infty^2$. This gives $r_{p,\rm max} \propto
v_\infty^{-4/7}$ (slower encounters can capture at larger separations).

We then sample the pericenter, $r_p$, uniformly in $[0, r_{p,\rm max}]$. This is
justified because in the gravitational focusing regime where $2GM / r_p \gg
v_\infty^2$, the cross-section for a capture with pericenter less than or equal to $r_p$
is proportional to $r_p$, so the probability density function for $r_p$ is
uniform. 

Finally, we compute $p(e | M, \mu, r_p)$ deterministically. This is done at the 
detector frame frequency of $10\,\mathrm{Hz}$ using Peters' equations \cite{peters1964gravitational}. 
Since GW captures have initial eccentricity $\sim 1$, the Peters constant reduces to $C_0 \approx 1.76\, r_p$.
We obtain the semi-major axis $a_{\rm ref}$ at 10 Hz from Kepler's third law and
invert
\begin{equation}
	\label{eq:peters}
	a_{\rm ref} = C_0 \, \frac{e_{\rm ref}^{12/19}\,(1 + 121\,e_{\rm ref}^2/304)^{870/2299}}{1 - e_{\rm ref}^2}
\end{equation}
to solve for $e_{\rm ref}$, yielding the eccentricity directly from $r_p$ without numerically integrating ODEs\footnote{For consistency
we also post-process using the method described in \cite{Vijaykumar:2024piy}
though this does not change the final constraint on $\sigma$ significantly.}.

\begin{figure}
\centering
\includegraphics[width=0.45\textwidth]{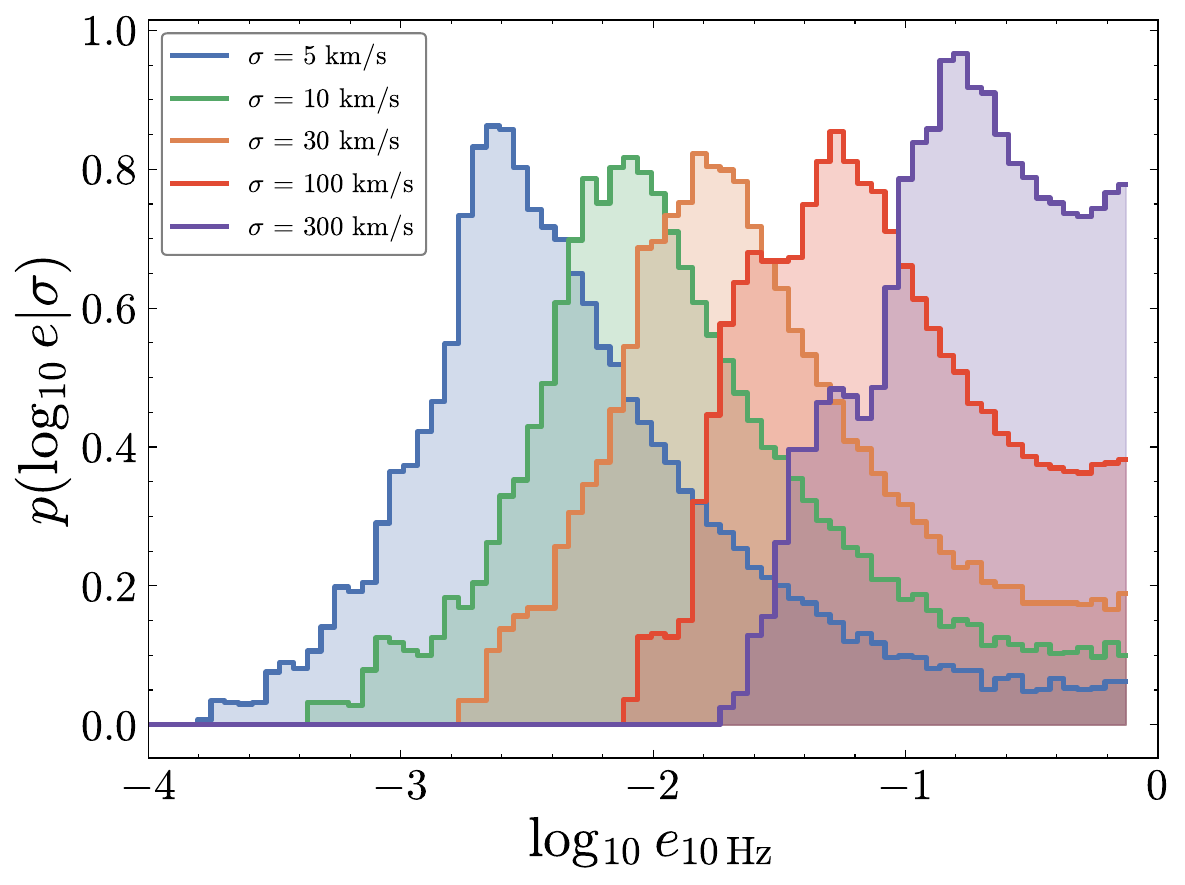}
\caption{Eccentricity distributions $p(\log_{10}                                                                                        
	e_{10\,\mathrm{Hz}} | \sigma)$ for different velocity
	dispersions, computed using the single-single GW capture                                                                                           
	model described in Section \ref{sec:eccentricity_distribution}. For this plot we use fiducial masses $m_1 = 30\,M_\odot$ and $m_2 =
	10\,M_\odot$. Higher velocity dispersions produce systematically higher
	eccentricities at fixed reference frequency. Crucially, the $e>0.05$
	tail---where LVK detectors can resolve eccentricity---shifts with $\sigma$. So
	while eccentricities below 0.05 are indistinguishable at LVK sensitivities,
	constraints in the $e > 0.05$ region serve
	as a proxy for the dispersion velocity.
	\label{fig:eccentricity_distribution}}
\end{figure}

The resulting eccentricity distribution forms our forward model,
$p(e | \sigma, \mu, M)$, shown in 
Fig.~\ref{fig:eccentricity_distribution}. It exhibits a peak at low eccentricities and a
tail toward high eccentricities that, for sufficiently large dispersions,
develops into a second peak. This structure can be understood from the
single-velocity distribution $p(e | v_\infty, \mu, M)$. 
The change of variables,
$p(e | v_\infty, \mu, M) = p(r_p | v_\infty, \mu, M)\,|dr_p/de|$, and the fact that
$p(r_p | v_\infty, \mu, M)$ is uniform on $[0,\, r_{p, \rm max}]$ implies that
the peak of $p(e | v_\infty, \mu, M)$ is set by the maximum of $|dr_p/de|$,
which always occurs at $r_p = r_{p,\rm max}$. The $r_{p,\rm max}$ is
determined by $v_\infty$, and a lower value of $v_\infty$ moves the peak to
smaller eccentricities. This explains the low eccentricity peak. The second peak
follows because for fixed $v_\infty$ the skewness of $p(e | v_\infty,
\mu, M)$ is always toward high eccentricities, never toward low: captures with
$r_p \ll r_{p,\rm max}$ are always kinematically accessible, so every $p(e |
v_\infty, \mu, M)$ has a tail extending to $e \sim 1$. These high-eccentricity
tails add coherently when marginalizing over $v_\infty$, producing the prominent
high-eccentricity feature in the full distribution. \textit{It is this second peak, not the tall first peak,
which distinguishes clusters with different dispersions at current LVK sensitivities}. This is because while the individual event 
eccentricity posterior below $0.05$ is flat, above $0.05$ we have constraining power. This property 
will be utilized in Section~\ref{sec:astro_implications}.

\section{Selection effects for eccentric BBHs}
\label{sec:selection_function}

At this stage, we could use the parameter estimation from Section~\ref{sec:results}, 
and eccentric population model from Section~\ref{sec:eccentricity_distribution}
to infer the dispersion velocity of clusters. However, 
if we directly use the results from Section~\ref{sec:results},
we would be assuming that these BBH mergers are a representative sample of the
underlying BBH merger population. In reality, this is not the case as we do not detect the majority of 
the BBH population due to Malmquist bias. In this section, we build the 
tools to account for this Malmquist bias in the context of eccentric BBHs.

Hierarchical inference accounts for Malmquist bias through
the population-level selection function
\begin{equation}
    \label{eq:selection_function}
    \alpha(\bm{\lambda}) = \int d\bm{\vartheta}\, p_{\text{det}}(\bm{\vartheta})\, p(\bm{\vartheta}|\bm{\lambda}),
\end{equation}
which gives the fraction of sources detectable under a population distribution
$p(\bm{\vartheta}|\bm{\lambda})$, with $p_{\text{det}}(\bm{\vartheta})$ the
per-source detection probability \citep{Mandel:2018mve, LIGOScientific:2021psn}. This 
selection function is then put into the denominator of the population likelihood (see Eq. \ref{eq:pop_likelihood}).

For eccentricity, $\alpha(\bm{\lambda})$ is particularly important because the
matched-filtering searches that produced the O4a catalog use quasicircular (QC)
template banks \citep{LIGOScientific:2021psn, LIGOScientific:2023lpe}, which can
miss $\sim 20\%$ of signals with $0.4 < e_{\text{gw, 10Hz}} < 0.6$
\citep{Gadre:2024ndy, Bhaumik:2024cec}. Note that coherent burst searches can mitigate this since they have higher
sensitivity to eccentric sources \citep{Das:2024zib}.

We compute $p_{\text{det}}(\bm{\vartheta})$ semi-analytically building upon
\citet{Gerosa:2024isl} and \citet{Essick:2023toz}. 
A similar semi-analytic treatment of selection effects for out-of-bank signals was recently explored in \citet{Vijaykumar:2026pqb} also 
with applications to eccentricity.
For Gaussian noise, the
matched-filter SNR \textit{maximized} over a QC template bank in detector $j$
approximately follows
\begin{equation}
    \label{eq:p_rho_mf}
    p(\rho^*_{\text{mf}, j} | \bm{\vartheta}) \approx \mathcal{N}\!\left(\rho_{\text{eff, opt}, j}(\bm{\vartheta}),\, 1\right),
\end{equation}
where $\rho_{\text{eff, opt}, j}$ is the noise-free overlap between the
eccentric signal $h_1$ and the best-fit QC template $h_2$,
\begin{equation}
    \label{eq:rho_eff_opt}
    \rho_{\text{eff, opt}, j}(\bm{\vartheta}_1) = \max_{\bm{\vartheta}_2}
    \frac{\langle h_1(\bm{\vartheta}_1) | h_2(\bm{\vartheta}_2) \rangle}
         {\sqrt{\langle h_2(\bm{\vartheta}_2) | h_2(\bm{\vartheta}_2) \rangle}}.
\end{equation}

The approximate equality in Eq.~(\ref{eq:p_rho_mf}) is due to two reasons.
First, the finite size of the template bank. If we include this effect it leads
to corrections to $\rho_{\text{eff, opt, j}}$. In particular 
\begin{equation}
    \label{eq:approx_inequality_reason_1}
    \rho^*_{\text{eff, opt, j}}(\bm{\vartheta}_1) \rightarrow 
    \rho^*_{\text{eff, opt, j}}(1 - \Delta \bm{\vartheta}_2^T \bm{\Gamma} \Delta \bm{\vartheta}_2)
\end{equation}
Where $\bm{\Gamma}$ is the Fisher information matrix and $\Delta \bm{\vartheta}^*_2 = \bm{\vartheta}^*_2 - \bm{\vartheta}^*_{2, \text{true}}$ is 
the deviation from the best fit parameters assuming a finite template bank to the best 
fit parameters assuming an infinitely dense template bank.

Second, the presence of noise. The template that maximizes the noise-free match 
is not generally the same as the one that maximizes the match in the presence of noise. Mathematically,
\begin{equation}
    \label{eq:approx_inequality_reason_2}
    \max_{\bm{\vartheta}_2} \langle h_1(\bm{\vartheta}_1) | h_2(\bm{\vartheta}_2) \rangle
    \neq
    \max_{\bm{\vartheta}_2} \left( \langle h_1(\bm{\vartheta}_1) | h_2(\bm{\vartheta}_2) \rangle + \langle n | h_2(\bm{\vartheta}_2) \rangle \right).
\end{equation}
This becomes especially important for signals where the SNR is low.
Ref.~\cite{Essick:2023toz} investigated this point in detail and 
found that Eq.~\ref{eq:p_rho_mf} is still a good approximation as long as the SNR threshold 
for detection is set sufficiently high. 
However, in comparison to Ref.~\cite{Essick:2023toz} our template bank differs from the signal model.
The ``best'' template often
has only modest overlap with the true signal, and templates in widely
separated regions of parameter space can have comparable overlaps. This
increases the effective covariance between filter outputs, so noise fluctuations
play a larger role in determining the maximized SNR. Intuitively, this should
make the approximation in Eq.~\ref{eq:p_rho_mf} less accurate. We 
investigate this in Appendix~\ref{sec:appendix_survival_function} and find that a threshold SNR 
of 10 is sufficient for this study. 

If we assume that the power is concentrated in the $(\ell, m) = (2, 2)$ mode and note we are using aligned-spin 
models, we can separate the contributions
of the intrinsic ($\bm{\theta}_i$) and extrinsic ($\bm{\phi}_i$) parameters. In particular, Eq.~(\ref{eq:rho_eff_opt}) becomes

\begin{equation}
    \label{eq:rho_eff_opt_extrinsic}
    \begin{aligned}
        \rho_{\text{eff, opt, j}}(&\bm{\theta}_1, \bm{\phi}_1) = \\
        &  w(\bm{\phi}_1) \max_{\bm{\theta}_2, t_{\text{2, coal}}, \phi_{\text{2, coal}}} 
        \frac{\langle h_{1 \times}(\bm{\theta}_1) | h_{2 \times}(\bm{\theta}_2, t_{\text{2, coal}}, \phi_{\text{2, coal}}) \rangle}
        {\sqrt{\langle h_{2 \times}(\bm{\theta}_2) | h_{2 \times}(\bm{\theta}_2) \rangle}}
    \end{aligned}
\end{equation}
where $w(\bm{\phi_1})$ depends on $(\iota_1, \alpha_1, \delta_1, \psi_1)$ and is given in Eq.~(3.31) in \cite{Finn:1992xs} (see also Eq.~(A7) in \cite{Gerosa:2020pgy}). 

We construct a stochastic QC
template bank using \texttt{SEOBNRv5\_ROM} \citep{Pompili:2023tna} at O4
sensitivity with the \texttt{pyCBC} bank-construction tools
\citep{Babak:2008rb, Harry:2009ea, Ajith:2012mn, Privitera:2013xza, Harry:2013tca, Capano:2016dsf}, and pre-select candidate templates using the
eccentric chirp-mass coordinate of \citet{Favata:2021vhw} to keep the
maximization tractable. The remaining maximization is implemented in
\texttt{JAX} \citep{jax2018github} with analytic optimization over $t_{\text{coal}}$
and $\phi_{\text{coal}}$, allowing us to perform the entire injection campaign in hours. 

With multiple detectors, the SNR adds in quadrature since the likelihood scales as $\exp(\sum_i \rho_i^2/2)$.
Therefore, the maximum network matched-filter SNR, $\rho^*_{\text{net, mf}}$, follows a non-central $\chi$ distribution with 
$k$ (number of detectors) degrees of freedom and a non-centrality parameter $\rho_{\text{eff, opt}}^2(\bm{\vartheta}_1) = \sum_j \rho^2_{\text{eff, opt, j}}(\bm{\vartheta}_1)$ (see Eq.~1.6 
in Ref.~\cite{ccbb70c4-d275-376a-8835-b08e66e59901}). One can then integrate $p(\rho^*_{\text{net, mf}} | \bm{\vartheta}_1)$ from a threshold SNR, $\rho_{\text{th}}$, to infinity and marginalize over $\bm{\phi}_2$
to obtain,
\begin{equation}
    \label{eq:p_rho_mf_det}
    p_{\text{det}}(\bm{\theta}_1) = \int d \bm{\phi}_1 \: Q_{k/2} \left( \rho_{\text{eff, opt}}(\bm{\theta}_1, \bm{\phi}_1) , \rho_{\text{th}} \right) p(\bm{\phi}_1),
\end{equation}
where $Q$ is the Marcum Q-function \cite{marcum1950table}.

\begin{figure}
    \includegraphics[width=0.5\textwidth]{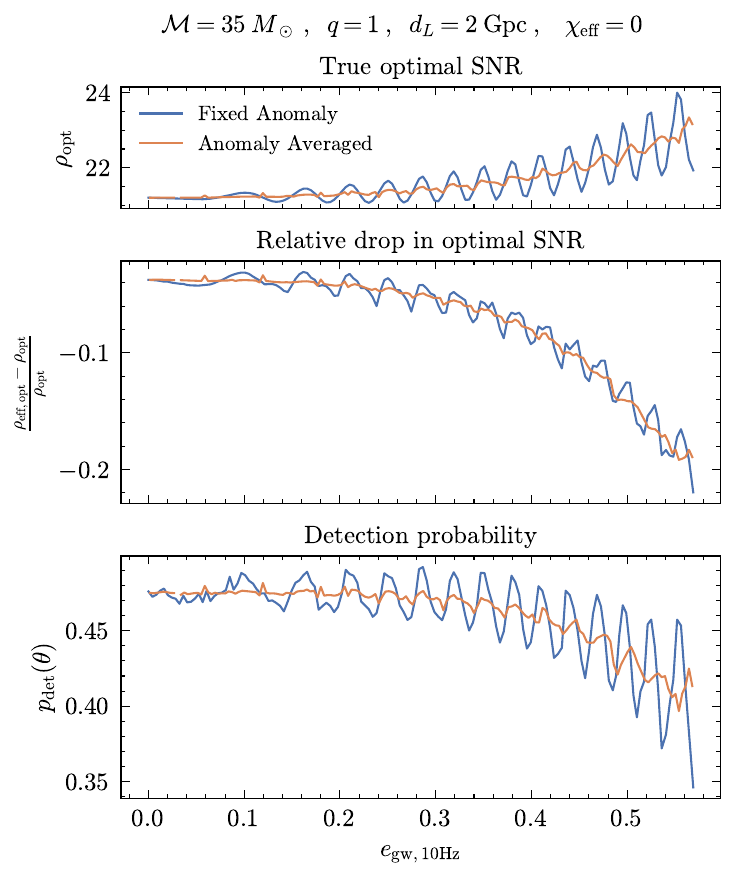}
    \caption{\label{fig:selection_function}
    True optimal SNR for a single detector (top), fractional SNR drop when
    maximizing over a quasicircular template bank (middle), and detection
    probability for an HLV network (bottom) as a function of $e_{\text{gw, 10Hz}}$
    for a fiducial BBH with $\mathcal{M} = 35\,M_\odot$, $q = 1$,
    $\chi_{\text{eff}} = 0$, and $d_L = 2\,\mathrm{Gpc}$. We use a 
	threshold SNR for detection of 10. Without averaging over
    the relativistic anomaly, the optimal SNR exhibits a periodic dependence on
    eccentricity that propagates into the SNR-loss and $p_{\text{det}}$ curves.
    Averaging over the anomaly (orange) suppresses this periodicity.}
\end{figure}

In Fig.~\ref{fig:selection_function} we show how $p_{\text{det}}(\bm{\theta})$
changes with eccentricity for a system with $\mathcal{M} = 35 M_\odot$, $q=1$,
$d_L = 2$Gpc and $\chi_{\text{eff}} = 0$ assuming Hanford, Livingston and Virgo (HLV) are 
online. The true optimal SNR has a periodic dependence on
eccentricity when using an orbit averaged frequency which causes
periodicity in $p_{\text{det}}(\bm{\theta})$. This is because the radial phase
parameter of eccentric binaries during plunge is not fixed, and therefore the
net energy emission is periodic \cite{Nee:2025zdy}. If one instead averages over
a uniform grid in the initial relativistic anomaly\footnote{One could also 
average over the mean anomaly which would result in the orbit averaged 
$p_{\text{det}}(\bm{\theta}).$}, the periodicity is
suppressed leading to the smoother orange curve
in Fig.~\ref{fig:selection_function}.

In practice we need to compute Eq. \ref{eq:selection_function} for many different
realizations of $\bm{\lambda}$. Therefore, we use a fiducial distribution
$p_\text{draw}(\bm{\vartheta})$ to write Eq.~\ref{eq:selection_function} as
a Monte Carlo estimate,
\begin{equation}
    \label{eq:selection_monte_carlo}
    \alpha(\bm{\lambda}) \approx \frac{1}{N_{\text{total}}} \sum_{\bm{\vartheta}_i \sim p_\text{draw}(\bm{\vartheta})}^{N_{\text{total}}} \frac{p_{\text{det}}(\bm{\vartheta}_i)\, p(\bm{\vartheta}_i | \bm{\lambda})}{p_\text{draw}(\bm{\vartheta}_i)},
\end{equation}
where $N_{\text{total}}$ is the total number of samples drawn from
$p_\text{draw}(\bm{\vartheta})$. The fiducial distribution
$p_\text{draw}(\bm{\vartheta})$ is chosen to be close to the true population so
as to minimize the Monte Carlo variance of this estimator
\citep{Farr:2019rap, Talbot:2023pex}.

To construct 
$p_\text{draw}(\bm{\vartheta})$ we use the same mass and redshift distributions 
as Ref.~\cite{Essick:2025zed} (see their Fig.
5). However, for the
component spin and eccentricity distribution we use aligned spins drawn independently from
a zero-mean truncated Gaussian, $\chi_1,\chi_2 \sim
\mathcal{N}{[-1,1]}(0,\sigma=0.5)$, in place of the precessing-spin distribution
of Ref.~\cite{Essick:2025zed}, and a log-uniform eccentricity distribution,
$\log_{10} e \sim \mathcal{U}(-4,\log_{10}0.8)$. We inject 40 million waveforms into Hanford and Livingston 
and use a hopeless SNR cut of 7 to filter the set of injections which are passed to the maximization algorithm pipeline (equation \ref{eq:rho_eff_opt}).
We use a threshold SNR of 10 as justified in appendix \ref{sec:appendix_survival_function}. 

\section{Astrophysical Implications}
\label{sec:astro_implications}

We now have posteriors from \ref{sec:results}, an eccentric population model 
from \ref{sec:eccentricity_distribution} and an eccentricity aware selection
function from \ref{sec:selection_function}. We are therefore ready to draw
astrophysical conclusions. While we do not find any individual events from O4a
that are confidently eccentric, we do obtain posterior distributions which
differ from the assumed priors.  For most events, this means excluding large
eccentricities. We can use these measurements to constrain the astrophysical
environments in which BBH mergers form. 

The two primary classes of dense stellar environments relevant for
dynamical BBH formation are globular clusters (GCs) and nuclear star clusters
(NSCs). GCs are ancient ($\gtrsim 10\,\mathrm{Gyr}$), gravitationally bound
stellar systems containing $\sim 10^5$--$10^6$ stars with typical masses of
$10^4$--$10^6\,M_\odot$ \citep{1996AJ....112.1487H, 2018MNRAS.478.1520B}.
Their central velocity dispersions span $\sigma \approx 1$--$20\,\mathrm{km/s}$
\citep{2018MNRAS.478.1520B, 2019MNRAS.482.5138B}.
NSCs are compact, massive stellar clusters found at the dynamical centers of
most galaxies with stellar masses $ \lesssim 10^{10}\,M_\odot$
\citep{2020A&ARv..28....4N, 2022ApJ...929...84B}. Typical NSC masses range from $10^6$--$10^8\,M_\odot$. NSCs exhibit substantially higher velocity
dispersions than GCs, with $\sigma \sim 20$--$200\,\mathrm{km/s}$ for most
systems, though the most massive can exceed $\sigma \gtrsim
200\,\mathrm{km/s}$ \citep{2017MNRAS.467.4180S, 2020A&ARv..28....4N, 2022ApJ...929...84B}.

Our analysis tests whether the data favors low-dispersion environments
consistent with GCs or high-dispersion environments consistent with NSCs.
Concretely, our null hypothesis is ``all BBHs observed by the LVK are explained
by single-single captures in nuclear star clusters''. We will then use the inferred 
dispersion to test this hypothesis. 

In order to consistently infer the population properties with 
selection effects, we need to also 
infer the mass, spin, and redshift population parameters which we denote
$\bm{M}_\text{pop}$, $\bm{S}_\text{pop}$, and $\bm{R}_{\text{pop}}$
respectively. This makes our full set of population parameters $\bm{\lambda} =
(\bm{M}_\text{pop}, \bm{S}_\text{pop}, \bm{R}_{\text{pop}}, \sigma)$. For
$\bm{M}_\text{pop}$ and $\bm{R}_{\text{pop}}$ we
adopt the fiducial \textsc{GWTC-4} population models of
\cite{LIGOScientific:2025pvj}: a two-peak, smoothed broken-power-law mass
distribution and a power-law redshift evolution of the merger rate. For the spin
distribution $\bm{S}_\text{pop}$, however, we cannot use the fiducial precessing
component-spin model, as \texttt{SEOBNRv5EHM} does not include precession; we therefore
model the effective spin $\chi_\text{eff}$ with a truncated Gaussian distribution,
$\bm{S}_\text{pop} = (\mu_{\chi_\text{eff}}, \sigma_{\chi_\text{eff}})$. Since
we do not have the full set of posterior samples and selection function computed
with \texttt{SEOBNRv5EHM} for O1--O3, we are restricted to inferring these population
parameters from O4a alone. 


Our full population posterior over $\bm{\lambda}$
is given by Bayes' rule,
\begin{equation}
\label{eq:full_pop_posterior}
p(\bm{\lambda} \,|\, \{d_i\}) \propto p(\bm{\lambda}) \prod_i p(d_i \,|\, \bm{\lambda}).
\end{equation}
We compute the per-event likelihood via Monte Carlo using the posterior samples $\{\bm{\vartheta}_{ij}\}_{j=1}^{N_s}$ for each event $i$:
\begin{equation}
	\label{eq:pop_likelihood}
	p(d_i \,|\, \bm{\lambda}) \propto \frac{1}{\alpha(\bm{\lambda})} \frac{1}{N_s} \sum_{j=1}^{N_s} \frac{p(\bm{\vartheta}_{ij} \,|\, \bm{\lambda})}{p(\bm{\vartheta}_{ij})},
\end{equation}
where $\alpha(\bm{\lambda})$ is the selection function and the $N_s$ posterior samples $\{\bm{\vartheta}_{ij}\}$ are drawn from a
density proportional to $p(d_i | \bm{\vartheta}_{ij})\, p(\bm{\vartheta}_{ij})$.
We sample the population posterior with \texttt{gwpopulation}
\citep{Talbot:2024yqw} using the \texttt{bilby} \citep{Ashton:2018jfp}
implementation of the \texttt{dynesty} \citep{Speagle:2019ivv} nested sampler,
with \texttt{nlive}$=512$ live points and the default \texttt{act-walk} sampling
scheme (\texttt{walks}$=100$, \texttt{nact}$=2$). We monitor the 
maximum uncertainty on the log-likelihood from Monte Carlo variance and ensure it does not
fall below $1.0$ following the recommendations from \citep{Talbot:2023pex}. 

\begin{figure*}[!t]
\centering
\includegraphics[width=\textwidth]{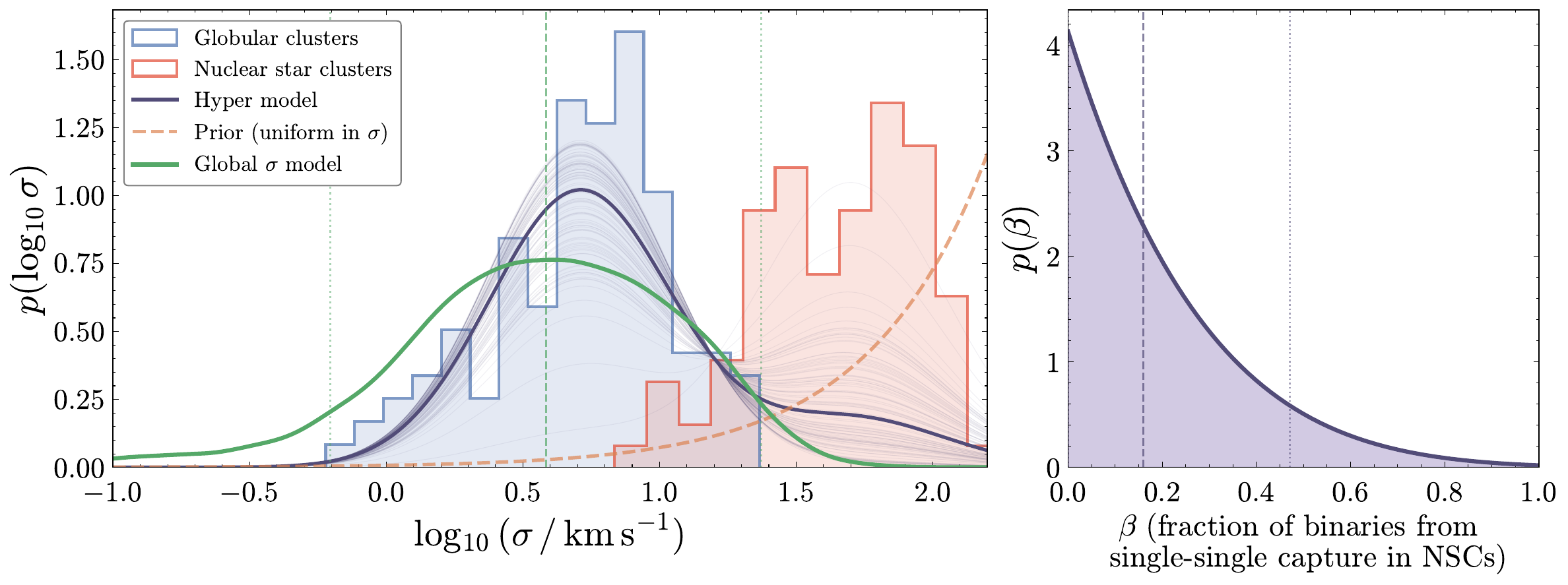}
\caption{
	Posterior constraints on the velocity dispersion $\sigma$ assuming all 56 O4a BBH
	mergers originate from single-single captures.
	\textit{Left:} the thick green curve shows
	the global $\sigma$ model posterior with median $3.9^{+19.7}_{-3.3}\,\mathrm{km/s}$ (90\%
	HDI). Vertical green dashed and dotted lines indicate its
	median and 90\% HDI. The dark purple curve shows the
	mean posterior predictive distribution (PPD) on $\sigma$ from the hyper model,
	which mixes GC and NSC populations with mixing fraction $\beta$, and the
	thin grey lines show individual PPD draws sampled from the $\beta$ posterior.
	The orange dashed curve shows the prior (uniform in $\sigma$). Blue and red
	histograms show the observed velocity dispersions of Milky Way globular
	clusters and NSCs, respectively. Log-normal fits to these
	histograms define the $p_{\rm GC}(\sigma)$ and $p_{\rm NSC}(\sigma)$
	components used in Eq.~(6).
	\textit{Right:} posterior on the NSC mixing fraction
	$\beta = 0.16_{-0.16}^{+0.31}$ (90\% HDI), with vertical dashed
	and dotted lines marking its median and 90\% HDI.
	Both models favor low velocity dispersions consistent with GC
	environments and disfavor the hypothesis that single-single captures in NSCs
	produce all observed BBH mergers in O4a.}
	\label{fig:sigma_posterior}
\end{figure*}

As opposed to the initial inference in Fig.~\ref{fig:bf_lineplot}, here
$p(e_{ij})$ is a log-uniform prior on $e_{\text{10Hz}}$, i.e., $p(e_{ij})
\propto 1/e_{ij}$, from $10^{-4}$ to $0.8$. This ensures support for low
eccentricities, which is important since the astrophysical forward model
$p(e|\sigma, M, \mu)$ has most of its support at low eccentricities for small
$\sigma$.  The single-single capture model does not generate eccentricities
below $10^{-4}$ for any reasonable value of $\sigma$ (see
Fig.~\ref{fig:sigma_posterior}, left panel), so the $10^{-4}$ lower bound
ensures sufficiently many usable samples after reweighting without biasing the
result. Setting the lower bound to $10^{-11}$ would place significant prior
density at eccentricities characteristic of isolated binaries and could cause
the constraint on $\sigma$ to become prior dominated in favor of GCs.  We
conservatively use a uniform prior on $\sigma$ from $0.1$ to
$1000\,\mathrm{km/s}$. For the mass, spin, and redshift
hyperparameters we adopt the fiducial population priors of the \textsc{GWTC-4}
populations analysis \citep{LIGOScientific:2025pvj}. We restate the full set of prior
distributions on all population hyperparameters in
Table~\ref{tab:pop_priors} in the appendix.

Here we are also only using 56 events as opposed to the 84 in Section~\ref{sec:results} since we are applying 
the event selection criteria that the SNR must be above 10. These SNRs are determined by 
looking at the maximum SNR recovered by the search pipelines in \cite{LIGOScientific:2025slb}. The list of 
events used for the hierarchical inference is displayed in Table \ref{tab:all_events}.

\begin{figure}
	\includegraphics[width=0.45\textwidth]{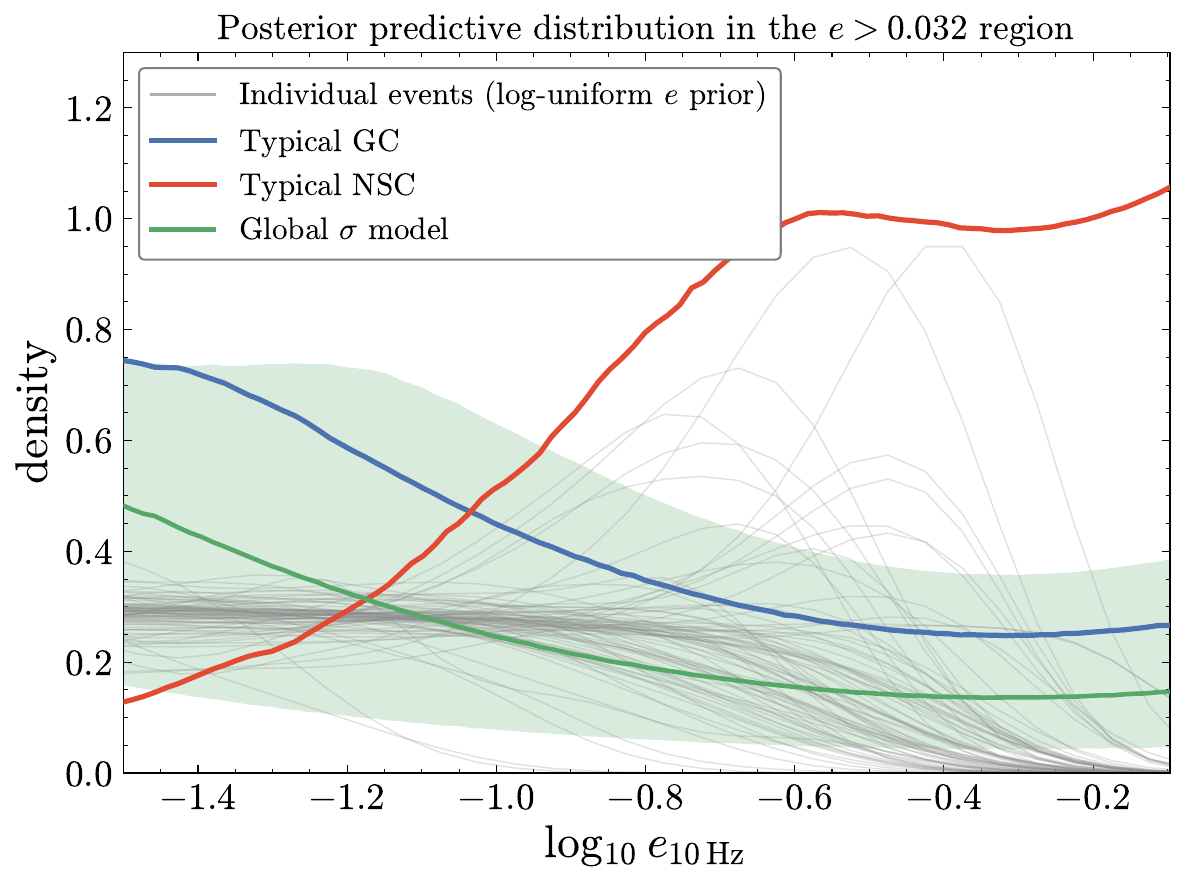}
	\caption{
		\label{fig:eccentricity_ppd}
		PPD (90 \% interval) of the eccentricity distribution under the global sigma model (shaded green).
		In red (blue) is the eccentricity distribution for a typical nuclear star cluster (globular cluster)
		assuming single-single capture.
		Despite the individual event parameter estimation and population models having support in $\log_{10} e_{\mathrm{10 Hz}} \in [-4, -0.1]$,
		we only display the tail of the eccentricity distribution. This is because below $e_{\mathrm{10 Hz}} \sim 0.05$ the parameter estimation is flat since
		there is very little likelihood change at low eccentricity. 
	}

\end{figure}

If we assume that all events share a common velocity dispersion $\sigma_{\rm global}$ 
then $\bm{\lambda} = (\sigma_{\rm global}, \bm{M}_\text{pop}, \bm{S}_\text{pop}, \bm{R}_{\text{pop}})$
and joint posterior is given by
\begin{equation}
p( \bm{\lambda} |\, \{d_i\}) \propto
p(\bm{\lambda})
\prod_i p(d_i \,|\, \bm{\lambda}).
\end{equation}
We report the marginal posterior over $\sigma_{\rm global}$
(Fig.~\ref{fig:sigma_posterior}, left panel)
and find $\sigma_{\rm global} < 19.7\,\mathrm{km/s}$ (95\% credible upper bound), with a median of $3.9\,\mathrm{km/s}$.  This result favors
low velocity dispersions consistent with GCs and disfavors NSCs as leading to
all BBHs via single-single captures. This is shown directly in
Fig.~\ref{fig:eccentricity_ppd}, where the predicted eccentricity distribution
under the global-$\sigma$ model lies well below that expected for a typical NSC.

In the second approach, we consider a hyper-model in which $\sigma$ is drawn from a
mixture of two populations calibrated to observed dense stellar environments. We
fit log-normal distributions to the central velocity dispersions of 112 Milky
Way GCs from \cite{2018MNRAS.478.1520B, 2019MNRAS.482.5138B} and 108 NSCs from
\cite{2022ApJ...929...84B}. The GC velocity dispersion profiles combine
ground-based radial velocities with \textit{Gaia} DR2 proper motions
\citep{2018A&A...616A..12G, 2019MNRAS.484.2832V}. The NSC 1D velocity
dispersions were computed by \cite{2017MNRAS.467.4180S} from King-profile
structural parameters measured by \cite{2004AJ....127..105B},
\cite{2006ApJS..165...57C}, and \cite{2014MNRAS.441.3570G}. The population-level
distribution of $\sigma$ is then modeled as
\begin{equation}
p(\sigma | \beta) = (1 - \beta)\, p_{\rm GC}(\sigma) + \beta\, p_{\rm NSC}(\sigma),
\end{equation}
where $\beta \in [0, 1]$ is the fraction of events originating from NSC-like environments. For each value of $\beta$, the per-event marginal likelihood is obtained by marginalizing over $\sigma$:
\begin{equation}
p(d_i | \beta) = \int p(d_i | \sigma)\, p(\sigma | \beta)\, d\sigma,
\end{equation}
and the posterior on $\beta$ follows from
\begin{equation}
p(\beta | \{d_i\}) \propto p(\beta) \prod_i p(d_i | \beta),
\end{equation}
with a uniform prior on $\beta$. The log-normal parameters for each component
are fixed from the empirical cluster catalogs and are not inferred. We find
$\beta = 0.16_{-0.16}^{+0.31}$ (90\% HDI, Fig.~\ref{fig:sigma_posterior}, right panel), consistent with
GC-dominated environments. Unless stated otherwise, point estimates are posterior
medians and uncertainties are 90\% HDIs computed with \texttt{arviz} \citep{arviz_2019};
the HDI on $\sigma$ is computed in $\log_{10}(\sigma)$ space (the space in which the
figure and the log-normal cluster models are defined) and then mapped back to
$\mathrm{km/s}$.

The above analysis was performed with only O4a events as analyzing O1-O3 events with 
consistent settings would require multiple new DINGO networks. Therefore,
to assess how the $\sigma$ constraint would respond to eccentric
events, as might be expected from O3 analyses \citep{Romero-Shaw:2022xko,Gupte:2024jfe}, we repeated the inference on a synthetic catalog
consisting of the O4a events supplemented with 3 injected eccentric events
(Gaussian posterior on $e$ with mean $0.3$ and standard deviation $0.065$) and
50 additional copies of O4a events with $\log_{10}\mathcal{B} < 0$.
The resulting $\sigma$ posterior shifts to lower dispersions, confirming that the
constraint is driven by the excess of events with no evidence for
eccentricity rather than the absence of confidently eccentric ones.

We emphasize that this analysis assumes the single-single capture model
exclusively; other dynamical formation channels would produce different
eccentricity distributions. The rates are therefore conditional on the BBHs
forming via single-single capture. Since we assume all BBHs are captures,
this maximizes the expected eccentric signal. The fact that even under
this ``favorable to capture'' assumption the data still disfavors
nuclear-cluster capture means the constraint is robust. A population with
only a sub-fraction of captures would produce even less eccentricity and further rule out
high $\sigma$. 

\section{Conclusions}
\label{sec:discussion}

In this paper, we present the analysis of 84 GW events from O4a, specifically targeting eccentricity. This is enabled
by the fast parameter estimation code \texttt{DINGO} which is aided by \texttt{Bilby} in 
cases where the sample efficiency is low or for regions of parameter space where we do not have trained networks. 
We investigate in further 
detail the 9 events which have the highest Bayes factors for eccentricity: GW230706\_104333, GW230709\_122727,
GW230820\_212515, GW231001\_140220, GW231114\_043211, GW231221\_135041,
GW231223\_032836, GW231224\_024321 and GW240104\_164932. 
When considering the astrophysical prior odds, none of
these events individually reaches sufficient statistical significance, as their odds ratios
do not exceed 1. Therefore, we conclude that there are no strong signs of eccentricity in the events observed in O4a. 

For the individual-event analysis in O4a, we find qualitative agreement in
the Bayes factors with \cite{Xu:2025xxx}, who employ the independent
\texttt{IMRPhenomTEHM} waveform model \cite{Planas:2025feq}. The events for which they report positive
support for eccentricity largely coincide with those in
Table~\ref{tab:interesting_table}.
The notable exceptions are GW230712\_090405 and GW231123\_135430, for
which they find much higher Bayes factors than we do.
The individual-event Bayes factors reported by
\cite{Zeeshan:2026pga, Malagon:2026uev} are systematically lower than ours, for
the nine events in Table~\ref{tab:interesting_table} they find
$\log_{10} \mathcal{B}_{\text{EAS/QCAS}} < 0$ in most cases
despite both analyses using the \texttt{SEOBNRv5EHM} model. 
It could be due to the choice of 
reference frequency, method to compute Bayes factors (we use a difference of evidence estimates whereas the Savage-Dickey is used in \cite{Zeeshan:2026pga, Malagon:2026uev}), sampling methodology
or sampling settings. This should be investigated in future work. 
We emphasize, that all three analyses agree on the central conclusion:
no O4a event shows substantial, statistically significant evidence for
eccentricity.

We also show the importance of marginalizing over the glitch in GW231114\_043211, GW231223\_032836,
and GW190701. In the case of GW190701, this causes a severe drop in the support for eccentricity 
from $\log_{10}	\mathcal{B}_{\text{EAS/QCAS}} = 3.68$ to $\log_{10}	\mathcal{B}_{\text{EAS/QCAS}} = 0.42$.
In future work, we plan to assess whether this drop persists when using an
eccentric signal model for the glitch mitigation.

Finally, by performing hierarchical inference under the assumption that all BBHs
form via single-single GW captures, we constrain the velocity
dispersion of the clusters hosting the observed systems to $\sigma = 3.9_{-3.3}^{+19.7}\,\mathrm{km/s}$ (90\% HDI). This disfavors single-single capture in nuclear star clusters ($\sigma
\sim 20-200\,\mathrm{km/s}$) as the origin of all observed BBH mergers in O4a, while
remaining consistent with single-single captures in globular cluster environments.

\appendix
\label{sec:appendix}

\section{2D distributions for O4a events}
\label{sec:appendix_posteriors}

\begin{figure*}
\includegraphics[width=\textwidth]{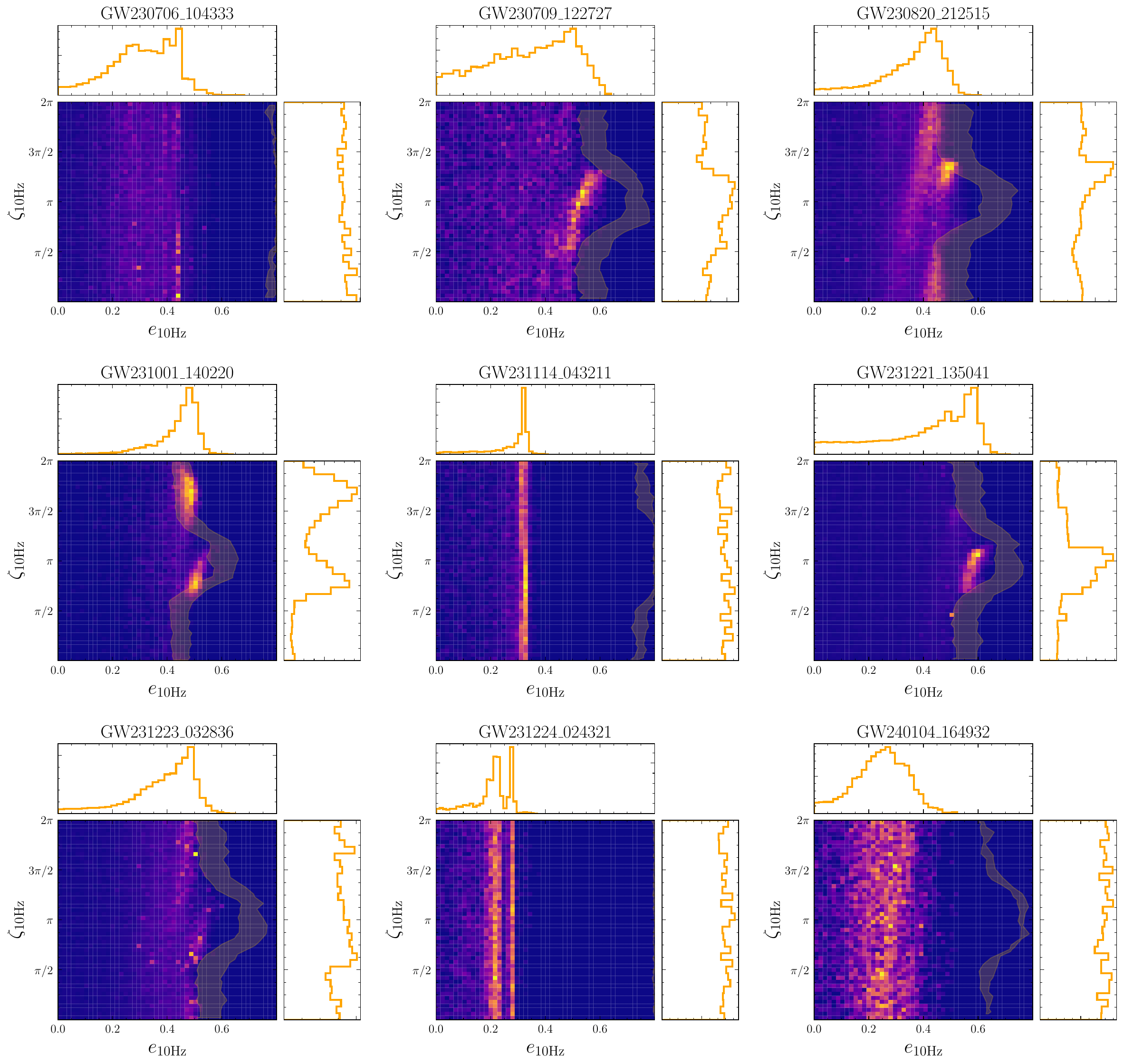}
\caption{\label{fig:posteriors} Posterior distributions of the
eccentricity, $e_{\text{10 Hz}}$ and relativistic anomaly $\zeta_{\text{10Hz}}$
of the 9 events reported in Table~\ref{tab:interesting_table}. Note $\zeta_{\text{10Hz}} = 0$
corresponds to periastron and $\zeta_{\text{10Hz}} = \pi$ corresponds to apastron.
Each posterior is colored by the density of samples with orange indicating a high density
and blue a low density. In the top and side panels for each event we show the 1D marginal distributions
for the eccentricity and relativistic anomaly. Colored in an orange band is the prior boundary
in $e_{\text{10Hz}}$ and $\zeta_{\text{10Hz}}$. This is a function of the total mass, which is
why the boundary is different for each event and a band as opposed to a line. Several events such
as GW230709\_122727, GW230820\_212515, GW231001\_140220 exhibit railing against the prior boundary.}
\end{figure*}

\section{Checking approximations in the selection function}
\label{sec:appendix_survival_function}

The purpose of this section is to check the approximation in
Eq.~\ref{eq:p_rho_mf}. This is an approximate equality as opposed to strict equality
because the best fit template is both a function of the
signal \textit{and} the noise as pointed out in \cite{Essick:2023toz}.

We recap the argument from \cite{Essick:2023toz} here. The whitened gravitational
wave data in frequency domain can be written as $d = h + n$, where $h$ is the whitened true signal
and $n$ is the noise realization. We denote the unit-normalized template bank as
$\{ \hat{h}_a \}$, where $\hat{h}_a = h_a / \sqrt{\langle h_a | h_a \rangle}$.
The matched filter output for template $a$ is
$\rho_{\rm mf, a} = \langle d | \hat{h}_a \rangle = \langle h | \hat{h}_a \rangle + \langle n | \hat{h}_a \rangle$.
This is distributed as $\mathcal{N}(\langle h | \hat{h}_a \rangle, 1)$
since $\langle n | \hat{h}_a \rangle \sim \mathcal{N}(0, 1)$ for stationary Gaussian noise.

But our detection statistic is $\rho_{\rm mf}^* = \max_a \rho_{\rm mf, a}$.
The survival function of this random variable can be written as
\begin{equation}
    \label{eq:survival_function}
    P(\rho^*_{\rm mf} > x) = P(\rho_{\rm mf, 1} > x \cup \rho_{\rm mf, 2} > x \cup \ldots).
\end{equation}
Since the noise contributions are correlated, the joint distribution of
$\{\rho_{\rm mf, a}\}$ is a non-identically distributed multivariate normal with non-trivial off-diagonal
covariance. In particular, $\text{Cov}[\rho_{\rm mf, a}, \rho_{\rm mf, b}] =
\langle \hat{h}_a | \hat{h}_b \rangle$.  Consequently, the CDF of $\rho^*_{\rm
mf}$ does not follow a normal distribution.
It also does not follow a Gumbel distribution since the variables are neither independently nor identically distributed.

Nevertheless, Ref.~\cite{Essick:2023toz} showed that
Eq.~\ref{eq:p_rho_mf} remains a good approximation for sufficiently high threshold SNR using a sine-Gaussian template model.
This is because in the high threshold SNR regime, only a small number of templates contribute appreciably to the maximum
and the correlation structure becomes less important.
However, since we are using a signal model which is different from the template bank,
the best template often may have only modest overlap with the true signal. Therefore,
widely separated regions of parameter space can have comparable overlaps and the
covariance between filter outputs increases, so noise fluctuations play a larger role in
determining the maximized SNR.

Therefore, we need to check the recommendations from Ref.~\cite{Essick:2023toz}.
Here we are interested in the noise-overlap properties of the maximization rather
than in the $(\ell, m) = (2, 2)$ mode approximation of the template bank, so we
fix the sky location and orientation to
$\alpha = 1, \delta = 0.5, \psi = 2.1, \theta_{\text{jn}} = 0, \phi_{\text{c}} = 0,
t_{\text{c}} = 0$ and vary only the intrinsic parameters and luminosity distance,
drawing them from the fiducial prior described in step 1 below. The
injections are generated with \texttt{SEOBNRv5EHM} and matched filtered, in the
$20$--$1024$~Hz band at the O4a fiducial PSD, against the quasi-circular
\texttt{SEOBNRv5\_ROM} template bank described in
Section~\ref{sec:selection_function}, using a two-detector (H1, L1) network.

\begin{figure}
    \centering
	\includegraphics[width=\textwidth]{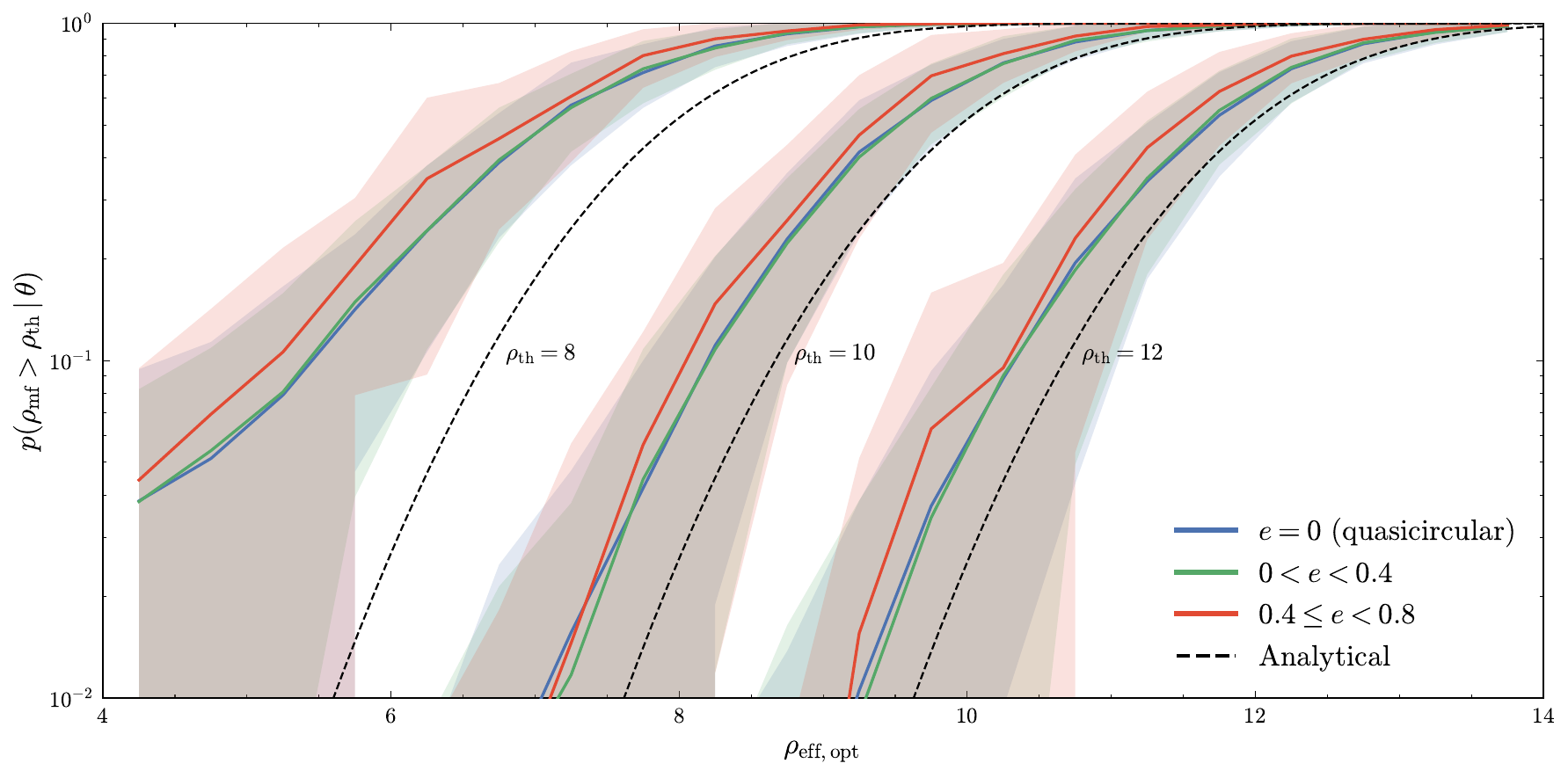}
    \caption{\label{fig:survival_function}
	  Probability of detecting injections assuming a quasi-circular template
      bank, comparing the semi-analytical approach (black dashed) against the
      empirically simulated survival function, for thresholds
      $\rho_{\text{th}} \in \{8, 10, 12\}$. The injections are stratified by
      eccentricity and overlaid: quasi-circular ($e=0$, blue), mildly eccentric
      ($0 < e < 0.4$, green) and strongly eccentric ($0.4 \le e < 0.8$, red). Each
      shaded band is the mean $\pm 2\sigma$ of the per-injection detection fraction (each
      measured over the 60 noise realizations) across the injections in a bin of
      $\rho_{\text{eff,opt}}$. 
	  For quasi-circular signals the best-matching template
      is a faithful match and the semi-analytical non-central $\chi^2$ reproduces the
      empirical survival function. As the eccentricity increases the quasi-circular
      templates match the signal only modestly, so widely separated regions of the
      bank attain comparable overlaps and the maximized SNR picks up larger noise
      fluctuations. This is why the high eccentricity (red) empirical survival function then lies increasingly above the
      semi-analytical curve compared to the quasi-circular case. This effect is most pronounced for lower $\rho_{\text{th}}$. }
\end{figure}

In Fig.~\ref{fig:survival_function} we compare the analytical survival
function from Eq.~\ref{eq:p_rho_mf_det} (blue), which assumes $\rho^*_{\rm mf}$ is normally
distributed per detector (Eq.~\ref{eq:p_rho_mf}) and therefore follows a non-central $\chi$
distribution with $k$ degrees of freedom for the network SNR ($k = 2$ for the H1--L1
network used here), against the true survival function
from Eq.~\ref{eq:survival_function} estimated empirically (red). The empirical
survival function is generated by

\begin{enumerate}
    \item drawing $\bm{\theta}_i$ from the fiducial prior and computing
    $\rho_{\text{eff, opt}, i}$ (Eq.~\ref{eq:rho_eff_opt_extrinsic}). The fiducial
    prior draws the source-frame chirp mass and mass ratio uniformly in the
    component masses over chirp masses of $15$--$50\,M_\odot$ and mass ratios of
    $0.125$--$1$; the eccentricity uniformly over $0$--$0.8$ (set to zero for the
    quasi-circular variant); the relativistic anomaly uniformly over a full cycle;
    the aligned component spins uniformly over $-0.99$ to $0.99$; and the luminosity
    distance uniformly over $3000$--$5000$~Mpc, with the sky location and
    orientation held fixed. We draw 3000 samples, evaluate
    $\rho_{\text{eff, opt}, i}$ for each, and retain a subset chosen to be
    approximately uniform in $\rho_{\text{eff, opt}}$ by stratified sampling in
    equal-width $\rho_{\text{eff, opt}}$ bins (2371 samples for the eccentric case
    and 2391 for the quasi-circular case);
    \item generating $n_{\text{noise}} = 60$ data realizations for each $\bm{\theta}_i$, i.e., $\bm{d}_{ij} = \bm{h}(\bm{\theta}_i) + \bm{n}_j$;
    \item finding the maximum network matched-filter SNR for each data realization, i.e., $\rho^*_{\text{mf}, ij} = \max_{\bm{\theta}_2} \langle \bm{d}_{ij} | \bm{h}(\bm{\theta}_2) \rangle$;
    \item estimating the empirical survival function for each $\bm{\theta}_i$ at a threshold $\rho_{\text{th}}$, i.e., $s_i(\rho_{\text{th}}) = p(\rho^*_{\text{mf}, i} > \rho_{\text{th}})$;
    \item plotting $\rho_{\text{eff, opt}, i}$ versus $s_i$ for $\rho_{\text{th}} \in \{8, 10, 12\}$.
\end{enumerate}
In step 4, $s_i(\rho_{\text{th}})$ is estimated by counting the fraction of the
$n_{\text{noise}}$ realizations with $\rho^*_{\text{mf}, ij} > \rho_{\text{th}}$.
Because $n_{\text{noise}} = 60$ is finite, we quantify the uncertainty on each
$s_i$ by bootstrapping over the noise realizations, and the red band shows the
bootstrap mean $\pm 1\sigma$. The band is broad both because of this finite-sample
bootstrap scatter and because the recovered SNR oscillates from sample to sample
across the fiducial draws at fixed $\rho_{\text{eff, opt}}$. Looking at the overlap of the empirical 
and semi-analytical curves in Fig.~\ref{fig:survival_function}
we conclude that $\rho_{\text{th}} = 10$ is sufficient for our study. 

\section{Table of events}
\label{sec:appendix_table}

\begin{center}
\begin{table*}[t]
\rowcolors{2}{gray!25}{white}
\setlength{\tabcolsep}{1pt}
\begin{minipage}{0.47\linewidth}
	\begin{footnotesize}
		\begin{tabular}{lCCCCCCC}
			Event name & $\text{FAR}_{\text{min}}$ ($\text{yr}^{-1}$) & $e_{10\text{Hz}}$ & $\log_{10} \mathcal{B}$ & $n_{\text{eff}}$ & $n_{\text{eff}}/N \times 100$ \\
			\hline
			GW230601\_224134$^{\dagger}$ & $< 1.0 \times 10^{-5}$ & $0.20_{-0.20}^{+0.19}$ & \ColorTwo$-0.32$ & \ColorTen$46091$ & \ColorSix$18.4\%$ \\ 
			GW230605\_065343$^{\dagger}$ & $< 1.0 \times 10^{-5}$ & $0.13_{-0.13}^{+0.12}$ & \ColorTwo$-0.48$ & \ColorThree$1311$ & \ColorOne$0.1\%$ \\ 
			GW230606\_004305$^{\dagger}$ & $1.3 \times 10^{-3}$ & $0.14_{-0.14}^{+0.16}$ & \ColorTwo$-0.59$ & \ColorTen$43038$ & \ColorFour$2.0\%$ \\ 
			GW230608\_205047$^{\dagger}$ & $1.2 \times 10^{-3}$ & $0.18_{-0.18}^{+0.17}$ & \ColorTwo$-0.25$ & - & - \\ 
			GW230609\_064958$^{\dagger}$ & $1.4 \times 10^{-4}$ & $0.18_{-0.18}^{+0.17}$ & \ColorTwo$-0.39$ & \ColorTen$54485$ & \ColorSeven$21.8\%$ \\ 
			GW230624\_113103$^{\dagger}$ & $1.8 \times 10^{-4}$ & $0.16_{-0.16}^{+0.15}$ & \ColorTwo$-0.46$ & \ColorTen$7719$ & \ColorFour$3.1\%$ \\ 
			GW230627\_015337$^{\dagger}$ & $< 1.0 \times 10^{-5}$ & $0.03_{-0.03}^{+0.03}$ & \ColorTwo$-0.90$ & - & - \\ 
			GW230628\_231200$^{\dagger}$ & $< 1.0 \times 10^{-5}$ & $0.13_{-0.13}^{+0.11}$ & \ColorTwo$-0.29$ & - & - \\ 
			GW230630\_125806 & $0.2$ & $0.21_{-0.21}^{+0.18}$ & \ColorTwo$-0.29$ & \ColorTen$63937$ & \ColorSeven$25.6\%$ \\ 
			GW230630\_234532 & $4.2 \times 10^{-4}$ & $0.13_{-0.13}^{+0.09}$ & \ColorTwo$-0.43$ & \ColorThree$3347$ & \ColorOne$0.2\%$ \\ 
			GW230702\_185453$^{\dagger}$ & $< 1.0 \times 10^{-5}$ & $0.12_{-0.12}^{+0.12}$ & \ColorTwo$-0.61$ & \ColorTen$12651$ & \ColorFive$5.1\%$ \\ 
			GW230704\_021211 & $0.2$ & $0.12_{-0.12}^{+0.12}$ & \ColorTwo$-0.62$ & \ColorTen$45613$ & \ColorSix$18.2\%$ \\ 
			GW230704\_212616 & $0.5$ & $0.22_{-0.21}^{+0.16}$ & \ColorThree$0.08$ & - & - \\ 
			GW230706\_104333 & $0.2$ & $0.31_{-0.19}^{+0.18}$ & \ColorFour$0.51$ & \ColorTen$8003$ & \ColorOne$0.3\%$ \\ 
			GW230707\_124047$^{\dagger}$ & $1.1 \times 10^{-3}$ & $0.17_{-0.17}^{+0.18}$ & \ColorTwo$-0.49$ & \ColorTen$70832$ & \ColorSeven$28.3\%$ \\ 
			GW230708\_053705 & $2.5$ & $0.22_{-0.22}^{+0.16}$ & \ColorThree$-0.18$ & \ColorTen$5496$ & \ColorFour$2.2\%$ \\ 
			GW230708\_230935$^{\dagger}$ & $3.7 \times 10^{-3}$ & $0.19_{-0.19}^{+0.19}$ & \ColorTwo$-0.41$ & \ColorTen$66651$ & \ColorSeven$26.7\%$ \\ 
			GW230709\_122727$^{\dagger}$ & $7.1 \times 10^{-2}$ & $0.32_{-0.27}^{+0.24}$ & \ColorFour$0.27$ & - & - \\ 
			GW230712\_090405 & $1.8 \times 10^{-2}$ & $0.14_{-0.14}^{+0.16}$ & \ColorTwo$-0.55$ & \ColorTen$27567$ & \ColorTwo$0.9\%$ \\ 
			GW230723\_101834$^{\dagger}$ & $3.4 \times 10^{-3}$ & $0.15_{-0.15}^{+0.13}$ & \ColorTwo$-0.42$ & \ColorTen$10760$ & \ColorFour$4.3\%$ \\ 
			GW230726\_002940$^{\dagger}$ & $< 1.0 \times 10^{-5}$ & $0.18_{-0.16}^{+0.14}$ & \ColorThree$-0.09$ & - & - \\ 
			GW230729\_082317 & $0.2$ & $0.19_{-0.19}^{+0.20}$ & \ColorTwo$-0.33$ & \ColorTen$7830$ & \ColorOne$0.4\%$ \\ 
			GW230731\_215307$^{\dagger}$ & $< 1.0 \times 10^{-5}$ & $0.09_{-0.09}^{+0.09}$ & \ColorTwo$-0.71$ & \ColorTen$5240$ & \ColorFour$2.1\%$ \\ 
			GW230803\_033412 & $3.0$ & $0.22_{-0.22}^{+0.19}$ & \ColorTwo$-0.26$ & \ColorTen$47946$ & \ColorSix$19.2\%$ \\ 
			GW230805\_034249 & $3.7 \times 10^{-3}$ & $0.16_{-0.16}^{+0.16}$ & \ColorTwo$-0.46$ & \ColorTen$8225$ & \ColorFour$3.3\%$ \\ 
			GW230806\_204041 & $3.7 \times 10^{-3}$ & $0.18_{-0.18}^{+0.17}$ & \ColorTwo$-0.42$ & \ColorTen$232958$ & \ColorNine$31.1\%$ \\ 
			GW230811\_032116$^{\dagger}$ & $< 1.0 \times 10^{-5}$ & $0.10_{-0.10}^{+0.11}$ & \ColorTwo$-0.68$ & \ColorTen$28303$ & \ColorSix$11.3\%$ \\ 
			GW230814\_061920$^{\dagger}$ & $6.3 \times 10^{-4}$ & $0.17_{-0.17}^{+0.17}$ & \ColorTwo$-0.45$ & \ColorTen$25359$ & \ColorFour$1.2\%$ \\ 
			GW230814\_230901$^{\dagger}$ & $< 1.0 \times 10^{-5}$ & $0.03_{-0.03}^{+0.03}$ & \ColorTwo$-0.94$ & - & - \\ 
			GW230819\_171910 & $0.011$ & $0.21_{-0.21}^{+0.19}$ & \ColorTwo$-0.31$ & \ColorTen$10239$ & \ColorFour$1.4\%$ \\ 
			GW230820\_212515 & $0.2$ & $0.35_{-0.19}^{+0.17}$ & \ColorFour$0.41$ & \ColorTen$28875$ & \ColorSix$11.6\%$ \\ 
			GW230824\_033047$^{\dagger}$ & $< 1.0 \times 10^{-5}$ & $0.16_{-0.16}^{+0.14}$ & \ColorTwo$-0.44$ & \ColorTen$83092$ & \ColorNine$33.2\%$ \\ 
			GW230825\_041334 & $0.1$ & $0.20_{-0.20}^{+0.17}$ & \ColorTwo$-0.31$ & \ColorTen$43905$ & \ColorSix$17.6\%$ \\ 
			GW230830\_064744$^{\dagger}$ & $12.0$ & $0.17_{-0.17}^{+0.14}$ & \ColorThree$-0.20$ & - & - \\ 
			GW230831\_015414 & $0.6$ & $0.29_{-0.29}^{+0.22}$ & \ColorThree$-0.10$ & \ColorTen$9730$ & \ColorFour$3.9\%$ \\ 
			GW230904\_051013$^{\dagger}$ & $3.9 \times 10^{-5}$ & $0.11_{-0.11}^{+0.08}$ & \ColorTwo$-0.58$ & \ColorTen$7678$ & \ColorOne$0.3\%$ \\ 
			GW230911\_195324$^{\dagger}$ & $1.4 \times 10^{-2}$ & $0.15_{-0.15}^{+0.14}$ & \ColorTwo$-0.38$ & - & - \\ 
			GW230914\_111401$^{\dagger}$ & $< 1.0 \times 10^{-5}$ & $0.09_{-0.09}^{+0.09}$ & \ColorTwo$-0.56$ & \ColorTen$24936$ & \ColorFive$10.0\%$ \\ 
			GW230919\_215712$^{\dagger}$ & $< 1.0 \times 10^{-5}$ & $0.09_{-0.09}^{+0.09}$ & \ColorTwo$-0.78$ & \ColorTen$14315$ & \ColorFive$5.7\%$ \\ 
			GW230920\_071124$^{\dagger}$ & $< 1.0 \times 10^{-5}$ & $0.24_{-0.24}^{+0.22}$ & \ColorTwo$-0.30$ & \ColorTen$6987$ & \ColorFour$2.8\%$ \\ 
			GW230922\_020344$^{\dagger}$ & $< 1.0 \times 10^{-5}$ & $0.19_{-0.17}^{+0.15}$ & \ColorThree$-0.10$ & \ColorTen$29590$ & \ColorSix$11.8\%$ \\ 
			GW230922\_040658$^{\dagger}$ & $< 1.0 \times 10^{-5}$ & $0.18_{-0.18}^{+0.14}$ & \ColorThree$0.14$ & - & - \\ 

		\end{tabular}
	\end{footnotesize}
\end{minipage}
\hspace{0.001\linewidth}
\begin{minipage}{0.47\linewidth}
	\rowcolors{2}{gray!25}{white}
	\begin{footnotesize}
		\begin{tabular}{lCCCCC}
			Event name & $\text{FAR}_{\text{min}}$ ($\text{yr}^{-1}$) & $e_{10\text{Hz}}$ & $\log_{10} \mathcal{B}$ & $n_{\text{eff}}$ & $n_{\text{eff}}/N \times 100$ \\
			\hline
			GW230924\_124453$^{\dagger}$ & $< 1.0 \times 10^{-5}$ & $0.10_{-0.10}^{+0.10}$ & \ColorTwo$-0.67$ & \ColorTen$10126$ & \ColorFour$4.1\%$ \\ 
				GW230927\_043729$^{\dagger}$ & $< 1.0 \times 10^{-5}$ & $0.10_{-0.10}^{+0.10}$ & \ColorTwo$-0.70$ & \ColorTen$75125$ & \ColorNine$30.0\%$ \\ 
			GW230927\_153832$^{\dagger}$ & $< 1.0 \times 10^{-5}$ & $0.05_{-0.05}^{+0.05}$ & \ColorOne$-1.00$ & \ColorTen$6908$ & \ColorFour$2.8\%$ \\ 
			GW230928\_215827$^{\dagger}$ & $1.5 \times 10^{-5}$ & $0.24_{-0.24}^{+0.20}$ & \ColorThree$-0.19$ & \ColorTen$18340$ & \ColorFive$7.3\%$ \\ 
			GW230930\_110730 & $0.2$ & $0.14_{-0.14}^{+0.14}$ & \ColorTwo$-0.55$ & \ColorTen$15604$ & \ColorFive$6.2\%$ \\ 
			GW231001\_140220$^{\dagger}$ & $1.6 \times 10^{-5}$ & $0.41_{-0.16}^{+0.14}$ & \ColorFive$1.07$ & - & - \\ 
			GW231004\_232346 & $0.2$ & $0.20_{-0.20}^{+0.20}$ & \ColorTwo$-0.36$ & \ColorTen$77932$ & \ColorNine$31.2\%$ \\ 
			GW231005\_021030$^{\dagger}$ & $1.0 \times 10^{-2}$ & $0.20_{-0.20}^{+0.17}$ & \ColorThree$0.00$ & - & - \\ 
			GW231005\_091549$^{\dagger}$ & $4.0 \times 10^{-2}$ & $0.14_{-0.14}^{+0.15}$ & \ColorTwo$-0.54$ & \ColorTen$29538$ & \ColorSix$11.8\%$ \\ 
			GW231008\_142521 & $1.6 \times 10^{-3}$ & $0.16_{-0.16}^{+0.17}$ & \ColorTwo$-0.49$ & \ColorTen$37494$ & \ColorSix$15.0\%$ \\ 
			GW231014\_040532 & $0.2$ & $0.16_{-0.16}^{+0.13}$ & \ColorTwo$-0.32$ & - & - \\ 
			GW231018\_233037 & $0.7$ & $0.16_{-0.14}^{+0.10}$ & \ColorThree$0.05$ & - & - \\ 
			GW231020\_142947$^{\dagger}$ & $< 1.0 \times 10^{-5}$ & $0.11_{-0.11}^{+0.09}$ & \ColorTwo$-0.25$ & - & - \\ 
			GW231028\_153006$^{\dagger}$ & $< 1.0 \times 10^{-5}$ & $0.13_{-0.13}^{+0.13}$ & \ColorThree$-0.21$ & - & - \\ 
			GW231029\_111508$^{\dagger}$ & $5.2 \times 10^{-5}$ & $0.13_{-0.13}^{+0.15}$ & \ColorTwo$-0.45$ & - & - \\ 
			GW231102\_071736$^{\dagger}$ & $< 1.0 \times 10^{-5}$ & $0.13_{-0.13}^{+0.12}$ & \ColorTwo$-0.56$ & \ColorTen$59031$ & \ColorSeven$23.6\%$ \\ 
			GW231104\_133418$^{\dagger}$ & $< 1.0 \times 10^{-5}$ & $0.13_{-0.13}^{+0.09}$ & \ColorTwo$-0.50$ & \ColorTen$11566$ & \ColorTwo$0.5\%$ \\ 
			GW231108\_125142$^{\dagger}$ & $< 1.0 \times 10^{-5}$ & $0.09_{-0.09}^{+0.09}$ & \ColorTwo$-0.76$ & \ColorTen$22077$ & \ColorFive$8.8\%$ \\ 
			GW231110\_040320$^{\dagger}$ & $< 1.0 \times 10^{-5}$ & $0.12_{-0.12}^{+0.14}$ & \ColorTwo$-0.66$ & \ColorTen$32486$ & \ColorFour$1.5\%$ \\ 
			GW231113\_122623 & $0.3$ & $0.20_{-0.20}^{+0.17}$ & \ColorTwo$-0.29$ & \ColorTen$6497$ & \ColorFour$2.6\%$ \\ 
			GW231113\_200417$^{\dagger}$ & $3.8 \times 10^{-5}$ & $0.15_{-0.15}^{+0.18}$ & \ColorTwo$-0.51$ & \ColorThree$4030$ & \ColorOne$0.2\%$ \\ 
			GW231114\_043211$^{\dagger}$ & $1.3 \times 10^{-4}$ & $0.27_{-0.16}^{+0.08}$ & \ColorFour$0.36$ & - & - \\ 
			GW231118\_005626$^{\dagger}$ & $< 1.0 \times 10^{-5}$ & $0.11_{-0.11}^{+0.11}$ & \ColorTwo$-0.31$ & - & - \\ 
			GW231118\_071402 & $2.8 \times 10^{-3}$ & $0.12_{-0.12}^{+0.13}$ & \ColorTwo$-0.61$ & \ColorTen$49385$ & \ColorSix$19.8\%$ \\ 
			GW231118\_090602$^{\dagger}$ & $< 1.0 \times 10^{-5}$ & $0.08_{-0.08}^{+0.08}$ & \ColorOne$-1.17$ & \ColorTen$22240$ & \ColorTwo$0.7\%$ \\ 
			GW231119\_075248 & $1.9 \times 10^{-2}$ & $0.18_{-0.18}^{+0.18}$ & \ColorTwo$-0.41$ & \ColorTen$102629$ & \ColorTen$41.1\%$ \\ 
			GW231123\_135430$^{\dagger}$ & $< 1.0 \times 10^{-5}$ & $0.18_{-0.17}^{+0.15}$ & \ColorThree$0.09$ & - & - \\ 
			GW231127\_165300 & $1.0 \times 10^{-2}$ & $0.17_{-0.17}^{+0.16}$ & \ColorTwo$-0.43$ & \ColorTen$59003$ & \ColorSeven$23.6\%$ \\ 
			GW231129\_081745 & $5.6 \times 10^{-2}$ & $0.16_{-0.16}^{+0.15}$ & \ColorTwo$-0.44$ & \ColorTen$26613$ & \ColorSix$10.6\%$ \\ 
			GW231206\_233134$^{\dagger}$ & $< 1.0 \times 10^{-5}$ & $0.13_{-0.13}^{+0.13}$ & \ColorTwo$-0.60$ & \ColorTen$37589$ & \ColorSix$15.0\%$ \\ 
			GW231206\_233901$^{\dagger}$ & $< 1.0 \times 10^{-5}$ & $0.10_{-0.10}^{+0.09}$ & \ColorTwo$-0.64$ & \ColorTen$16695$ & \ColorFive$6.7\%$ \\ 
			GW231213\_111417$^{\dagger}$ & $< 1.0 \times 10^{-5}$ & $0.20_{-0.19}^{+0.14}$ & \ColorThree$-0.11$ & \ColorTen$16907$ & \ColorFive$6.8\%$ \\ 
			GW231221\_135041$^{\dagger}$ & $0.5$ & $0.42_{-0.31}^{+0.21}$ & \ColorThree$0.21$ & \ColorTen$9446$ & \ColorFour$1.3\%$ \\ 
			GW231223\_032836$^{\dagger}$ & $3.8 \times 10^{-4}$ & $0.38_{-0.18}^{+0.18}$ & \ColorFour$0.53$ & \ColorThree$4175$ & \ColorFour$1.7\%$ \\ 
			GW231223\_075055 & $1.6$ & $0.10_{-0.10}^{+0.13}$ & \ColorTwo$-0.27$ & - & - \\ 
			GW231223\_202619$^{\dagger}$ & $2.0 \times 10^{-3}$ & $0.14_{-0.14}^{+0.13}$ & \ColorTwo$-0.34$ & - & - \\ 
			GW231224\_024321$^{\dagger}$ & $< 1.0 \times 10^{-5}$ & $0.20_{-0.11}^{+0.09}$ & \ColorFour$0.65$ & - & - \\ 
			GW231226\_101520$^{\dagger}$ & $< 1.0 \times 10^{-5}$ & $0.05_{-0.05}^{+0.05}$ & \ColorTwo$-0.91$ & \ColorTen$16101$ & \ColorTwo$0.7\%$ \\ 
			GW231230\_170116 & $0.4$ & $0.26_{-0.24}^{+0.19}$ & \ColorThree$-0.09$ & \ColorTen$89282$ & \ColorNine$35.7\%$ \\ 
			GW231231\_154016$^{\dagger}$ & $< 1.0 \times 10^{-5}$ & $0.40_{-0.39}^{+0.32}$ & \ColorThree$0.01$ & - & - \\ 
			GW240104\_164932$^{\dagger}$ & $< 1.0 \times 10^{-5}$ & $0.24_{-0.16}^{+0.16}$ & \ColorFour$0.48$ & - & - \\ 
			GW240107\_013215 & $2.8 \times 10^{-2}$ & $0.24_{-0.23}^{+0.17}$ & \ColorThree$-0.07$ & \ColorTen$19950$ & \ColorFive$8.0\%$ \\ 
		\end{tabular}
	\end{footnotesize}
\end{minipage}
\caption{\label{tab:all_events} Table of the 84 BBH events analyzed in
	this work. For each candidate, we report the minimum FAR
obtained by the search pipelines, the 90\% HDI of the inferred eccentricity at 10 Hz, the $\log_{10}$ Bayes factor
	comparing EAS and QCAS
	hypotheses, the effective sample size and the sample
	efficiency. Events which are sampled with \texttt{Bilby} instead of \texttt{DINGO} have dashes 
	for the effective sample size and sample efficiency. We usually sample with
	\texttt{Bilby} when we have changed the frequency range or do not have trained networks
	to accommodate the detector configuration. We also sample with \texttt{Bilby} for 8 events for which
	the variance of the weights is unbounded and we therefore cannot obtain a high number of effective samples. These
	are GW230627\_015337, GW231014\_040532, GW231018\_233037, GW231020\_142947, GW231114\_043211, GW231118\_005626, GW231223\_075055, and GW231224\_024321.
	Here, we only use glitch subtraction as opposed to glitch marginalization (reported in Table~\ref{tab:interesting_table}). The $\dagger$ superscript 
	on the events indicates that this was one of the 56 events used for population inference in Section~\ref{sec:astro_implications}.
	}
\end{table*}
\end{center}

%
%


\clearpage

\section{Consistency of mass, spin and redshift populations with GWTC-4}
\label{sec:hierarchical_posterior_on_other_params}

\begin{figure*}
	\includegraphics[width=\textwidth]{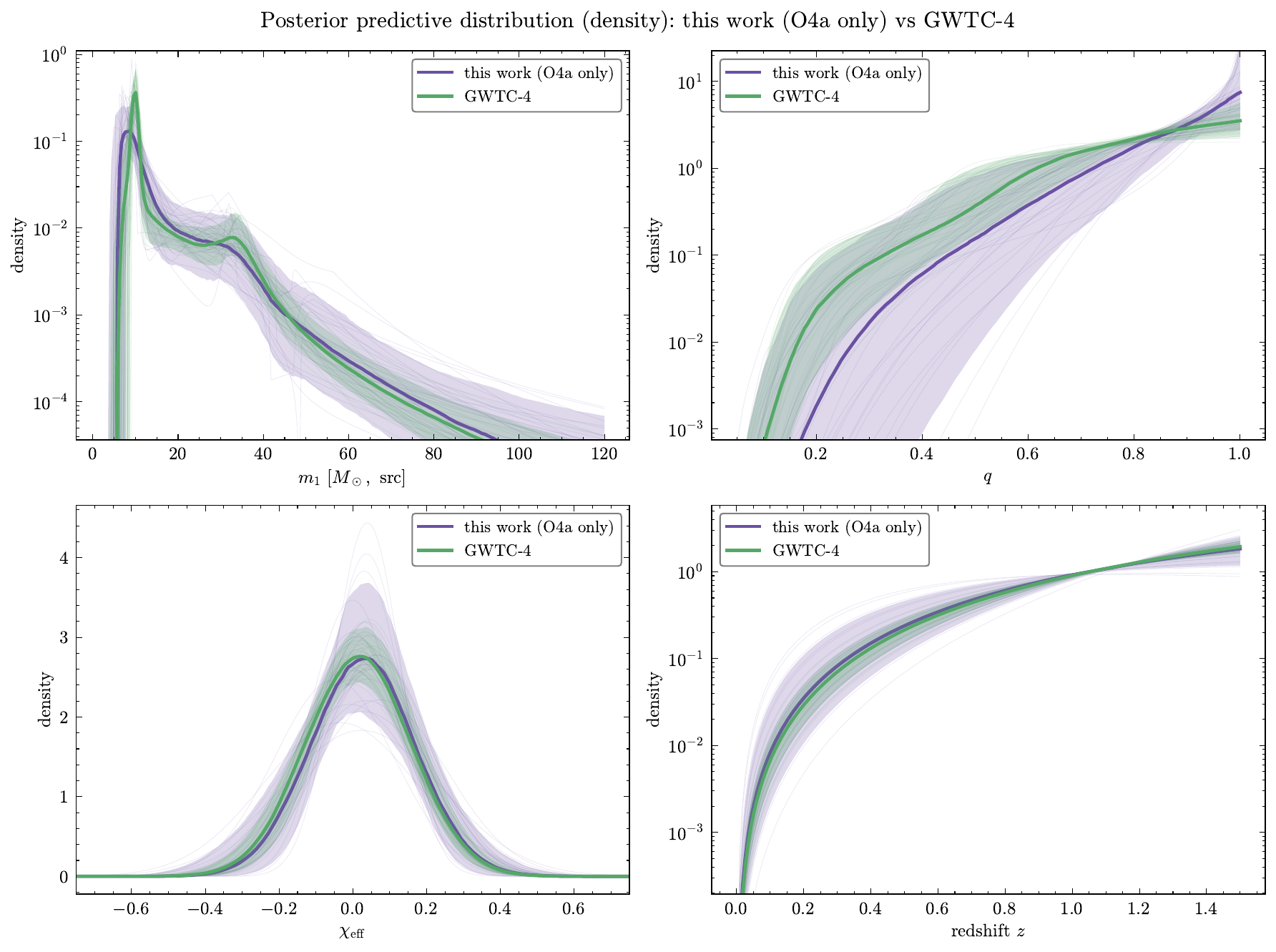}
	\caption{\label{fig:hyper_posterior_comparison} Posterior predictive distributions (PPDs) of the primary mass (top left), 
	mass ratio (top right), effective spin (bottom left) and redshift (bottom right) when jointly inferred with 
	the single-single capture eccentricity model described in Section \ref{sec:eccentricity_distribution}. In purple, we 
	plot the results from this work and in green we plot the results from \citep{LIGOScientific:2025pvj}. The shaded 
	region denotes the 90\% credible interval, the solid line denotes the median and the thin lines represent 
	individual draws from the PPD. We see general agreement with the LVK results, though our posteriors are broader since 
	we are analyzing events only from O4a whereas the LVK result includes events from O1-O4a. }
\end{figure*}

While the main result of the paper is the inference of the dispersion velocity due to eccentricity, 
we would like to check the consistency of our mass, redshift and spin population inference
with \citep{LIGOScientific:2025pvj}. We do this by comparing the posterior predictive distributions in Figs. \ref{fig:hyper_posterior_comparison}.
We find qualitative agreement with the LVK results. However, since we are only including events from O4a, our population posteriors are broader than the 
LVK results. Further work (with consistent settings) is needed to determine the effect of including 
an eccentric population model on the population properties of the mass, spin and redshift distributions. 

\begin{table*}[t]
\begin{center}
\footnotesize
\renewcommand{\arraystretch}{1.2}
\setlength{\tabcolsep}{4pt}
\begin{tabular}[t]{lll}
	\toprule
	Parameter & Description & Prior \\
	\midrule
	\multicolumn{3}{l}{\textit{Mass distribution} (\textsc{BrokenPowerLawTwoPeaks})} \\
	\midrule
	$\alpha_1$ & \pdesc{Low-mass power-law slope} & Uniform $[-4.0,\ 12.0]$ \\
	$\alpha_2$ & \pdesc{High-mass power-law slope} & Uniform $[-4.0,\ 12.0]$ \\
	$\beta_q$ & \pdesc{Mass-ratio power-law slope} & Uniform $[-4.0,\ 12.0]$ \\
	$m_{\rm break}$ & \pdesc{Primary-mass break point $[M_\odot]$} & Uniform $[5.0,\ 60.0]$ \\
	$\lambda_0$ & \pdesc{Mixture weight, power-law component} & Uniform $[0.0,\ 1.0]$ \\
	$\lambda_1$ & \pdesc{Mixture weight, first Gaussian peak} & Uniform $[0.0,\ 1.0]$ \\
	$\mu_1$ & \pdesc{Mean of first Gaussian peak $[M_\odot]$} & Uniform $[2.0,\ 20.0]$ \\
	$\mu_2$ & \pdesc{Mean of second Gaussian peak $[M_\odot]$} & Uniform $[20.0,\ 50.0]$ \\
	$\sigma_1$ & \pdesc{Width of first Gaussian peak $[M_\odot]$} & Uniform $[0.1,\ 10.0]$ \\
	$\sigma_2$ & \pdesc{Width of second Gaussian peak $[M_\odot]$} & Uniform $[0.4,\ 10.0]$ \\
	$m_{\rm low,1}$ & \pdesc{Low-mass cutoff, first component $[M_\odot]$} & Uniform $[2.0,\ 20.0]$ \\
	\bottomrule
\end{tabular}
\hspace{0.0cm}
\begin{tabular}[t]{lll}
	\toprule
	Parameter & Description & Prior \\
	\midrule
	\multicolumn{3}{l}{\textit{Mass distribution} (cont.)} \\
	\midrule
	$m_{\rm low,2}$ & \pdesc{Low-mass cutoff, second component $[M_\odot]$} & Uniform $[2.0,\ 20.0]$ \\
	$\delta_{m,1}$ & \pdesc{Low-mass smoothing scale, first $[M_\odot]$} & Uniform $[0.0,\ 10.0]$ \\
	$\delta_{m,2}$ & \pdesc{Low-mass smoothing scale, second $[M_\odot]$} & Uniform $[0.0,\ 10.0]$ \\
	$m_{\rm max}$ & \pdesc{Maximum mass $[M_\odot]$} & Fixed $300.0$ \\
	\midrule
	\multicolumn{3}{l}{\textit{Redshift evolution} (\textsc{PowerLawRedshift})} \\
	\midrule
	$\kappa$ & \pdesc{Merger-rate evolution index, $(1+z)^{\kappa}$} & Uniform $[-2.0,\ 8.0]$ \\
	\midrule
	\multicolumn{3}{l}{\textit{Spin distribution} (Gaussian in $\chi_{\rm eff}$)} \\
	\midrule
	$\mu_{\chi_{\rm eff}}$ & \pdesc{Mean of effective-spin distribution} & Uniform $[-1.0,\ 1.0]$ \\
	$\sigma_{\chi_{\rm eff}}$ & \pdesc{Width of effective-spin distribution} & Uniform $[0.01,\ 0.5]$ \\
	\midrule
	\multicolumn{3}{l}{\textit{Eccentricity} (single-single capture, Section~\ref{sec:eccentricity_distribution})} \\
	\midrule
	$\sigma$ & \pdesc{1D env.\ velocity dispersion $[\mathrm{km\,s^{-1}}]$} & Uniform $[0.0,\ 10^{2.5}]$ \\
	\bottomrule
\end{tabular}
\end{center}
\caption{\label{tab:pop_priors} Prior distributions on the population
hyperparameters of the four-component hierarchical model fit to the BBH events,
comprising a two-peak smoothed broken-power-law mass distribution
(\textsc{BrokenPowerLawTwoPeaks}), a power-law redshift evolution of the merger
rate (\textsc{PowerLawRedshift}; redshift support $z_{\rm max}=3.0$), a Gaussian
effective-spin distribution truncated to $\chi_{\rm eff}\in[-1,1]$, and a
single-single dynamical-capture eccentricity model
(Section~\ref{sec:eccentricity_distribution})
parametrized by the 1D environmental velocity dispersion $\sigma$. The
functional forms of the mass, spin, and redshift distributions follow the
\textsc{GWTC-4} populations analysis \citep{LIGOScientific:2025pvj}.
}
\end{table*}

\subsection{Projecting component-spin models into $\chi_{\rm eff}$}
\label{sec:chi_eff_projection}

The fiducial \textsc{GWTC-4} spin model from \citep{LIGOScientific:2025pvj} is
parametrized in terms of the component
spin magnitudes and tilts. The
magnitudes follow $a_1, a_2 \sim \mathrm{TruncNorm}(\mu_\chi, \sigma_\chi)$ on
$[0, a_{\max}]$, and the cosine tilts follow a two-component mixture, an aligned
truncated Gaussian with weight $\xi$ and an isotropic component otherwise. However, 
we model $\chi_{\text{eff}}$ and do not consider spin tilts since spin-precession is 
not modeled in \texttt{SEOBNRv5EHM}. Thus we cannot directly compare our spin distribution 
to \textsc{GWTC-4}. 

Therefore, we project each GWTC-4 hyperposterior sample in the $\chi_{\rm eff}$.
For each draw we generate a synthetic catalog
of $N = 2000$ events. The component masses $(m_1, m_2)$ are sampled from the mass
model evaluated at that draw's mass hyperparameters, the spin magnitudes
$a_1, a_2$ are drawn from $\mathrm{TruncNorm}(\mu_\chi, \sigma_\chi)$ on
$[0, a_{\max}]$ and the cosine tilts $\cos\theta_{1,2}$ are drawn from the
two-component mixture $\xi\,\mathrm{TruncNorm}(\mu_t, \sigma_t)$ plus
$(1-\xi)\,\mathcal{U}(-1,1)$. The effective spin of each synthetic event is
\begin{equation}
\label{eq:chi_eff_def}
\chi_{\rm eff} = \frac{m_1 a_1 \cos\theta_1 + m_2 a_2 \cos\theta_2}{m_1 + m_2},
\end{equation}
and we record the mean $\mu_{\chi_{\rm eff}}$ and standard deviation
$\sigma_{\chi_{\rm eff}}$ of the resulting $\chi_{\rm eff}$ sample. Repeating over
all hyperposterior draws yields the induced posterior on the
population-predictive $\chi_{\rm eff}$ moments, which we use to display the
$\chi_{\rm eff}$ marginal of the population-predictive distribution.
The masses are grid-sampled from the model
PDF on a fixed $(m_1, q)$ mesh.
This Monte-Carlo change of
variables makes the component-spin parametrization of the \textsc{GWTC-4} model
somewhat comparable to the truncated-Gaussian $\chi_{\rm eff}$ distribution we
infer from O4a.

\begin{acknowledgments}

We would like to thank Oliver Long and Matthew Mould for useful discussions 
during the preparation of this manuscript. 

The computational work for this manuscript was carried out
on the compute clusters Saraswati, Lakshmi and Hypatia at the Max
Planck Institute for Gravitational Physics in Potsdam.
\texttt{DINGO} is publicly available through the \texttt{Python} package
dingo-gw at \texttt{https://github.com/dingo-gw/dingo}. Stable
versions of \texttt{DINGO} are published through conda.
This work made use of the \texttt{Python} packages
\texttt{NumPy}~\citep{harris2020array},
\texttt{SciPy}~\citep{2020SciPy-NMeth},
\texttt{pandas}~\citep{mckinney-proc-scipy-2010, reback2020pandas}, and
\texttt{Matplotlib}~\citep{Hunter:2007}.
The authors also acknowledge the use of Anthropic's Claude~\citep{anthropic2024claude}
as a coding assistant. 

A.R.B. is supported by the Veni research programme which is (partly) financed by
the Dutch Research Council (NWO) under the grant VI.Veni.222.396; he
acknowledges support from the Spanish Agencia Estatal de Investigación grant
PID2022-138626NB-I00 funded by MICIU/AEI/10.13039/501100011033 and the ERDF/EU,
PID2024-157460NA-I00; and the Spanish Ministerio de Ciencia, Innovación y
Universidades (Beatriz Galindo, BG23/00056), co-financed by UIB. This work was
supported by the Universitat de les Illes Balears (UIB); the Spanish Agencia
Estatal de Investigación grants PID2022-138626NB-I00, RED2024-153978-E,
RED2024-153735-E, funded by MICIU/AEI/10.13039/501100011033 and the ERDF/EU; and
the Comunitat Autònoma de les Illes Balears through the Conselleria d'Educació i
Universitats with funds from the European Union - NextGenerationEU/PRTR-C17.I1
(SINCO2022/6719) and from the European Union - European Regional Development
Fund (ERDF) (SINCO2022/18146). 

A.K. thanks the International Max Planck Research School for Intelligent Systems (IMPRS-IS) for support. S.R.G. and L.P. are supported by a UKRI Future Leaders Fellowship (grant number MR/Y018060/1). A.B.'s research is supported in part by the European Research Council (ERC) Horizon Synergy Grant “Making Sense of the Unexpected in the Gravitational-Wave Sky” grant agreement no. GWSky–101167314.
This research has made use of data or software obtained from the Gravitational
Wave Open Science Center (gwosc.org), a service of the LIGO Scientific
Collaboration, the Virgo Collaboration, and KAGRA. This material is based upon
work supported by NSF's LIGO Laboratory which is a major facility fully funded
by the National Science Foundation, as well as the Science and Technology
Facilities Council (STFC) of the United Kingdom, the Max-Planck-Society (MPS),
and the State of Niedersachsen/Germany for support of the construction of
Advanced LIGO and construction and operation of the GEO600 detector. Additional
support for Advanced LIGO was provided by the Australian Research Council. Virgo
is funded, through the European Gravitational Observatory (EGO), by the French
Centre National de Recherche Scientifique (CNRS), the Italian Istituto Nazionale
di Fisica Nucleare (INFN) and the Dutch Nikhef, with contributions by
institutions from Belgium, Germany, Greece, Hungary, Ireland, Japan, Monaco,
Poland, Portugal, Spain. KAGRA is supported by Ministry of Education, Culture,
Sports, Science and Technology (MEXT), Japan Society for the Promotion of
Science (JSPS) in Japan; National Research Foundation (NRF) and Ministry of
Science and ICT (MSIT) in Korea; Academia Sinica (AS) and National Science and
Technology Council (NSTC) in Taiwan.

\end{acknowledgments}

\bibliographystyle{aasjournalv7}
\bibliography{references}

\end{document}